\documentclass[12pt]{iopart}
\usepackage[pdftex]{graphicx} 

\usepackage{amsmath,amssymb,color,hyperref,graphicx,pdfsync,overpic,color,epstopdf,rotating,dashrule,float,bm,multicol,setspace,cancel}
\usepackage{lipsum}
\usepackage[table]{xcolor}
\usepackage{tabularx,booktabs}
\usepackage{mathrsfs}
\usepackage{mathtools}
\usepackage[numbers]{natbib}
\usepackage{epsfig,epstopdf}
\usepackage{subcaption} 

\usepackage{xcolor}

\definecolor{green}{rgb}{0.3960, 0.6040, 0.1180}

\begin{document}

\title{Extracting dominant dynamics about unsteady base flows}

\author{Alec J. Linot, Barbara Lopez-Doriga, Yonghong Zhong, and Kunihiko Taira}

\address{Department of Mechanical and Aerospace Engineering, University of California, Los Angeles, CA 90095, USA}
\ead{alinot5@g.ucla.edu} 
\vspace{10pt}
\begin{indented}
\item[]June 2025
\end{indented}

\begin{abstract}

A wide range of techniques exist for extracting the dominant flow dynamics and features about steady, or periodic base flows. However, there have been limited efforts in extracting the dominant dynamics about unsteady, aperiodic base flow. These flows appear in many applications such as when there is a sudden change in the flow rate through a pipe, when an airfoil experiences stall, or when a vortex forms. For these unsteady flows, it is valuable to know not only the dynamics of the base flow but also the features that form around this base flow. 
Here, we discuss the current state of research on extracting important flow structures and their dynamics in such cases with time-varying base flows. In particular, we consider data-driven decompositions, operator-based methods, causality analysis, and some other approaches.  We also offer an outlook and call attention to key areas that require future efforts.
\end{abstract}

% ------------------------------ % 
\section{Introduction}
\label{sec:intro}

Unsteady flows exhibit rich dynamics and spatial patterns. Modern simulations and experiments can capture these unsteady flow fields with remarkable spatial and temporal resolutions, but the underlying dynamics may still be difficult to interpret.
Instead of attempting to track every flow structure and its temporal evolution, it is beneficial to extract the dominant low-dimensional dynamics that play an important role in the overall behavior of the system. 
In fact, our research community has been successful in extracting such dominant flow structures, for which modal analysis has served as a foundation. Data-driven and operator-based modal analysis techniques~\cite{HLBR-11,berkooz1993pod,kutz2016book,Schmid2022,taira2017modal} have been widely used for a variety of research efforts \cite{taira2020modal}. Particularly noteworthy are the proper orthogonal decomposition~\cite{lumley1967structure,aubry1988dynamics}, dynamic mode decomposition~\cite{schmid_2010}, stability analysis~\cite{Schmid2001, Drazin81, Theofilis:ARFM11, Joslin09}, and resolvent analysis~\cite{Trefethen1993,jovanovic2005componentwise,mckeon2010resolvent}.  

\begin{figure}
    \centering
    \includegraphics[width=0.9\textwidth]{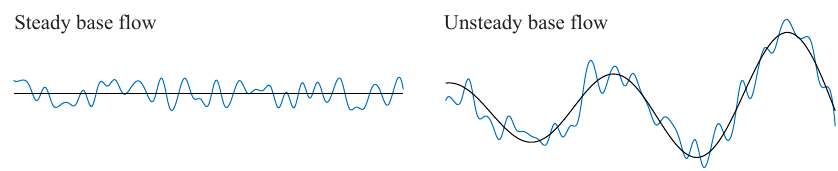}  
    \caption{Perturbation dynamics about (left) steady and (right) unsteady base flows.  Base flows are shown in black and the perturbations on top of the base flow are shown in blue.}
    \label{fig:setup}
\end{figure}

With these techniques, the research community has made great strides in analyzing fluid flows with steady base states. However, the examination of complex flow dynamics with respect to time-varying base states, as shown in figure \ref{fig:setup}, remains a challenge. The difficulty stems from the lack of theories and analysis techniques available for time-varying base states, as well as the guidelines on how to choose the time-varying base states themselves. Floquet analysis is one method for analyzing perturbation dynamics about time-periodic base flows \cite{Floquet:1883, Perko:2013, JordanSmith:1999}, but this method does not readily extend to aperiodic time-varying base flows.  

In fact, many fluid-based systems possess unsteady base flows, whose acceleration and deceleration significantly influence the overall flow dynamics. For example, notable acceleration and deceleration occur in both internal flows, such as pipe flow \citep{GREENBLATT2004}, and external flows around moving bodies, such as maneuvering airfoils \citep{SENGUPTA2007}. Even when there does not exist a clear transient force driving the system, transient periods of acceleration or deceleration are observed in many examples, including in turbulent channel flows \cite{moin1990boundary3d,lozanoduran2021boundary3d}, mixing in a tank \cite{dagadu2015mixing,zhou2022cfd}, dynamic stall \cite{McCroskey1981}, and unpredictable weather patterns \cite{krishnamurthy2019predictability,kautz2022atmospheric}.

Here, we call attention to the shortcomings of the standard techniques established for steady base flows when applied to time-varying base flows. We also present existing methods that can extract dominant, low-dimensional flow dynamics and spatial flow structures with respect to time-varying flows. Analysis techniques for unsteady base flows are far less prevalent than the methods for steady base flows listed above. We hope that calling attention to these approaches will expand their use and motivate the development of new methods.
Also presented in this paper are the challenges associated with studying unsteady flows, including the determination of the unsteady base flow, examination of the time-varying flow evolution operator, and the greater computational requirements compared to analysis for flows with steady base states.

In what follows, we discuss our perspective on recent approaches for analyzing unsteady flows. First, we outline some fundamental challenges with applying analysis techniques to flows with unsteady base states and with finding time-varying base flows in section~\ref{sec:preliminaries}. Then, we present data-driven modal approaches in section~\ref{sec:approaches_modal}, operator-based approaches (nonmodal stability analysis, optimally time-dependent modes analysis, and input-output analysis) in sections~\ref{sec:approaches_nonmodal}-\ref{sec:approaches_resolvent}, causality analysis in section~\ref{sec:approaches_causality}, and other approaches in section~\ref{sec:approaches_other}. Finally, we provide concluding remarks in section~\ref{sec:discussion}. We hope that this work stimulates future discussions and the development of computationally efficient techniques for the analysis of statistically evolving flows in time.

\section{Preliminaries}
\label{sec:preliminaries}

In this paper, we aim to understand the evolution of perturbations (fluctuations) around some unsteady base flow (state). These perturbations may be analyzed using both linear and nonlinear approaches. If the perturbation is sufficiently small, linear methods are a valid approach.  However, once the magnitudes of these perturbations become large, nonlinear effects need to be taken into account.  In this section, we present the equations of motion for the perturbation about an unsteady base flow to highlight both the linear and nonlinear effects driving the perturbation dynamics.  We will also describe why simple extensions of traditional linear methods are insufficient for understanding unsteady flows.

\subsection{Governing equations}
Let us consider an $n$-dimensional state vector $\mathbf{q}(t)\in\mathbb{R}^n$ of some dynamical system
\begin{equation}
\label{eq:general}
    \dfrac{d\mathbf{q}(t)}{dt} = \mathbf{N}(\mathbf{q}(t)) + \mathbf{f}(t),
\end{equation}
where $\mathbf{N}:\mathbb{R}^n\rightarrow\mathbb{R}^n$ is a nonlinear function and $\mathbf{f}(t)\in\mathbb{R}^n$ is a time-varying forcing term. For the systems of interest in this paper, \eqref{eq:general} comes from discretizing a partial differential equation for a field variable (or vector of field variables) $q(\mathbf{x},t)$ onto a finite grid, or a finite set of basis functions, to give the time-dependent vector $\mathbf{q}(t)$. Although, we write the state vector as a real-valued vector for simplicity, there should be a similar formulation if the state vector is complex, for example, from a spatial Fourier transform. 

Next, we express the state variable and forcing term to be comprised of their base variables $[\bar{\mathbf{q}}(t),\bar{\mathbf{f}}(t)]$ and perturbations  $[\mathbf{q}'(t),\mathbf{f}'(t)]$, such that 
\begin{equation}
    \mathbf{q}(t)=\bar{\mathbf{q}}(t)+\mathbf{q}'(t)
\end{equation}
and
\begin{equation}
    \mathbf{f}(t)=\bar{\mathbf{f}}(t)+\mathbf{f}'(t).
\end{equation}
By inserting these expressions into \eqref{eq:general} and performing a Taylor series expansion of $\mathbf{N}$ about $\bar{\mathbf{q}}(t)$, we find that
\begin{equation}
\label{eq:generalpert}
    \dfrac{d\mathbf{q}'(t)}{dt} =\mathbf{L}[\bar{\mathbf{q}}(t)]\mathbf{q}'(t) + \check{\mathbf{f}}(t),
\end{equation}
where $\mathbf{L}[\bar{\mathbf{q}}(t)]=\nabla\mathbf{N}|_{\bar{\mathbf{q}}(t)}$ and
\begin{equation}
\label{eq:generalforce}
    \check{\mathbf{f}}(t) = -\dfrac{d\bar{\mathbf{q}}(t)}{dt}+\mathbf{N}[\bar{\mathbf{q}}(t)] + \bar{\mathbf{f}}(t)+\mathbf{f}'(t)+\mathcal{O}(||\mathbf{q}'(t)||^2)
\end{equation}
collects all of the remaining terms. This formulation frames the evolution of perturbations in a time-varying, forced linear system. 
While the above equation may suggest a simple application of linear dynamical systems theory around some chosen base flow to understand the perturbation dynamics, care is needed in adopting this formulation.

Different treatments of the forcing term will lead to different interpretations of the results. These different treatments correspond to choices of the base states $[\bar{\mathbf{q}}(t),\bar{\mathbf{f}}(t)]$ and the characteristics of the perturbations $[\mathbf{q}'(t),\mathbf{f}'(t)]$. If the base states are the solution to the Navier--Stokes equations (NSE), the first three terms of \eqref{eq:generalforce} sum to zero.  In this case, $\check{\mathbf{f}}(t)$ reduces to $\mathbf{f}'(t)+\mathcal{O}(||\mathbf{q}'(t)||^2)$. Moreover, if the magnitude of $\mathbf{q}'(t)$ is small, then the higher-order terms can be neglected.

Another option is to choose a time-varying base flow that is not a solution to the NSE, such as a temporally filtered or ensemble-averaged flow.  In this case, $\check{\mathbf{f}}(t)$ retains all terms.  
This assumption does not require that the perturbations, or the deviations of the base flow from the NSE, are small. However, this perspective embeds these effects as part of the forcing term, which should be carefully analyzed as the nonlinearity can play a role as an internal feedback mechanism in the overall dynamics.

When the base flow is time-invariant, a solution of the NSE, and there is no forcing, standard linear stability analysis shows the asymptotic behavior of perturbations. For a diagonalizable constant matrix, we can write 
\begin{equation} \label{eq:StabStationary}
    \mathbf{L}=\mathbf{V}\Lambda\mathbf{V}^{-1}.
\end{equation}
Combining \eqref{eq:generalpert} and \eqref{eq:StabStationary} results in the equations in the eigenvector coordinate system
\begin{equation} 
    \dfrac{d\mathbf{y}}{dt}=\Lambda\mathbf{y},
\end{equation}
where $\mathbf{y}=\mathbf{V}^{-1}\mathbf{q}'$. This is a set of decoupled equations with the solution 
\begin{equation}
    \mathbf{y}=e^{\Lambda t}\mathbf{c}
\end{equation}
or
\begin{equation} \label{eq:EigSimple}
    y_k=c_k e^{\lambda_k t}=c_k e^{\lambda_{R,k} t}e^{\text{i}\lambda_{I,k} t} 
    = c_k e^{\lambda_{R,k} t} [ \cos(\lambda_{I,k} t) + \text{i} \sin(\lambda_{I,k} t) ],
\end{equation}
where $\lambda_k=\lambda_{R,k}+\text{i}\lambda_{I,k}$ and $\mathbf{c}$ depends on the initial condition. From \eqref{eq:EigSimple} we see that the real part of the eigenvalue is associated with exponential growth or decay, and the imaginary part of the eigenvalue is associated with oscillations (due to Euler's formula). Often, we worry about the eigenvector with the largest real eigenvalue because the perturbation will either grow most rapidly or decay most slowly along this direction. When we are only interested in the asymptotic behavior, standard linear stability analysis is sufficient, but this analysis is insufficient for understanding transient perturbation dynamics about unsteady base flows.

When the base flow is time-varying, both the eigenvalues and eigenvectors of $\mathbf{L}[\bar{\mathbf{q}}(t)]$ evolve in time. Standard linear stability analysis, or the eigendecomposition, of $\mathbf{L}[\bar{\mathbf{q}}(t)]$ provides us with the `stability' of an instantaneous profile as if it were `frozen,' which is known as a quasi-steady approach \citep{Shen1961}. The system's eigenvalues provide the asymptotic dynamics of the perturbation, which makes this approach valid only when the base flow is time-invariant or evolves much slower than the perturbation. One attempt to overcome this challenge involves time-averaging the flow over some time window and performing stability analysis about this mean \cite{Smyth_Peltier_1994}. However, this introduces the problem of the time-averaged flow not necessarily corresponding to a physically realizable flow field.  As such, stability analysis would not be appropriate when the base flow varies on a time scale similar to those of the perturbations.  

Another approach is to determine the instantaneous growth of perturbations, as opposed to finding the asymptotic dynamics. This is possible using the energy method \cite{Serrin1959,Joseph1976}, which determines the critical Reynolds number where a perturbation will lead to immediate energy growth (i.e., $d||\mathbf{q}'||/dt>0$). The time-varying nature of the flow does not matter here because this analysis provides instantaneous growth. In \cite{Conrad1965} this method was applied to accelerating and decelerating flows. The drawbacks of this approach are that it only provides a lower bound on the critical Reynolds number, and it gives no indication as to the dynamics of the perturbation. This lower bound estimate tends to be much lower than the critical Reynolds number observed in experiments \citep{Serrin1959}. Instead of these approaches, we will later discuss ones that elucidate the transient dynamics -- not just the instantaneous or asymptotic behavior.  

\subsection{Time-varying base flows}
\label{sec:baseflow}

Determining the base flow is important in how we interpret the forcing in \eqref{eq:generalforce}. First, we give background on typical approaches to finding stationary base flows. Then, we present methods for finding unsteady base flows, including analytical methods, numerical methods, time-averaging methods for turbulent flows, and exact coherent structures methods.

There are a few ways to determine time-invariant (steady) base states.  For analysis that requires access to the steady solution to the Navier--Stokes equations, either a known exact solution or a numerical solution can be used.  For a stable steady state, a numerical simulation can be performed over a sufficiently long time until all unsteadiness in the flow subsides and the flow converges to the stable steady base flow.  On the other hand, if the flow of interest is unstable, the unstable steady state needs to be determined through a Newton-Raphson solver \citep{Kawahara2011} or the selective frequency damping method \cite{aakervik2006sfd}.  Determining unstable steady states can be difficult, especially for high Reynolds number conditions.

In some problem setups, the choice of time-averaged flow as the time-invariant base state can be useful.  While time-averaged flows are not generally a solution to the NSE, their use as a base flow in some analyses can be appropriate. If the time-averaged flow is not a solution, then the first three terms on the right-hand side of \eqref{eq:generalforce} do not cancel out, and there will be a nonzero forcing $\check{\mathbf{f}}(t)$.  As linear stability analysis typically assumes the forcing to be zero, caution should be taken when performing linear stability analysis with the mean flow as the base state~\cite{beneddine2017unsteady}.  Since linearization about a time-averaged state does not strictly follow dynamical systems theory, an analysis based on the mean flow should be considered as a modeling effort \cite{taira2017modal}. 
However, resolvent analysis can serve as a powerful technique to analyze turbulent flows about their mean, given that their fluctuations are statistically stationary in time.  This assumption enables the effects of the fluctuations to be modeled as external harmonic forcing input to the flow system~\cite{mckeon2010resolvent}.

For cases with time-varying base flows, the ideal situation would be to have direct access to a known time-varying solution to the NSE.
Although limited in numbers, analytical time-varying solutions are available for some basic flows. Some notable examples include Stokes' first and second problems \citep{Schlichting2017,Batchelor2000,Liu2008}, pulsatile flow in a channel \citep{Majdalani2008} and pipe \citep{Womersley1955}, and pipes and ducts with arbitrary pressure gradients \citep{Fan1964,Urbanowicz2023}. More recently, Linot et al.\ \cite{Linot2024} found a generalized parallel channel flow solution with time-varying wall motion and pressure gradient. These are situations where the base flow varies in one coordinate direction and time, making solutions tractable. Unfortunately, the problem becomes more challenging when the solution varies in two or three dimensions. For these flows, matched asymptotic expansions can, in some cases, offer an analytical solution \citep{VanDyke1975}. One such solution exists for flow past an impulsively started cylinder \citep{Bar-Lev_Yang_1975} (also see \cite{Chien:1977} for corrections), although its accuracy and validity should be carefully considered. 

When analytical solutions are unavailable, numerical solutions or experiments of time-varying base flows may be used in some scenarios. Whether or not to consider a numerical solution or experiment as the time-varying base flow depends upon the stability of the problem. In these cases, we only have direct access to the full flow $\mathbf {q}(t)$. Thus, we can only get the base flow if the perturbation $\mathbf{q}'(t)$ rapidly decays to 0, such that $\mathbf {q}(t)\approx\bar{\mathbf q}(t)$. Even if the base flow is an attracting trajectory about which $\|{\mathbf q}’(t)\| \rightarrow 0$ as $t \rightarrow \infty$, the perturbation needs to decay quickly such that $\bar{\mathbf q}(t)$ can be found over the time window of interest.  For dynamics that are not asymptotically stable $\|{\mathbf q}’(t)\| \nrightarrow 0$ as $t \rightarrow \infty$, ${\mathbf q}’(t)$ will exist as a non-zero time-varying variable, which complicates the separation process.  In this latter case, it may be possible to assume ${\mathbf q}’(t)$ to possess certain statistical properties (frequency content), assuming that the time scales of $\bar{\mathbf q}(t)$ and ${\mathbf q}’(t)$ do not overlap.  

When the flow is turbulent ($\|{\mathbf q}’(t)\| \nrightarrow 0$ as $t \rightarrow \infty$), and we do not have any information on statistical properties ${\mathbf q}’(t)$, we need other approaches to approximate the base flow $\bar{\mathbf q}(t)$.
One possibility is to approximate the base flow using a filtering procedure.  
In such a case, the application of a low-pass filter or a moving temporal average on the full flow field $\mathbf q(t)$ may be utilized in an effort to remove $\mathbf q’(t)$.  An ensemble average is also an option if we can access multiple cases of the events.  There is also the possibility to use the triple decomposition that separates the flow field into the mean, coherent, and random components~\cite{reynolds1972mechanics, Edstrand_2016}. 
In such a case, the sum of the mean and coherent components may be used as $\bar{\mathbf q}(t)$.  

Another option when selecting the base flow is to directly compute some unstable solution of the Navier--Stokes equations.
These solutions are known as exact coherent states, or structures, (ECS) \cite{Kawahara2011,Graham2021}, and are typically equilibria, traveling waves, periodic orbits, or relative periodic orbits. Many ECS have been discovered in plane Couette \cite{Nagata1990,Waleffe1998,Waleffe2001,Kawahara2001,Gibson2008,Viswanath2007,Linot2023}, plane Poiseuille flow \citep{Waleffe2003,Park2015,Shekar2018}, and pipe flow \citep{Faisst2003,Wedin2004} using Newton-Raphson methods. Fixed points have also been discovered for flow around a cylinder using selective frequency damping \cite{aakervik2006sfd}. These ECS organize turbulence by attracting and repelling trajectories. Using these ECS, we can define the base flow as either the ECS or the heteroclinic connections linking them \citep{Halcrow2009}. The ECS solutions can be treated as stationary or periodic, which allows us to use standard linear stability or Floquet analysis. However, the heteroclinic connections between ECS are transient and require the methods we will describe to analyze. These heteroclinic connections represent prototypical paths a turbulent trajectory takes through state space, making their analysis an important -- and largely unexplored -- final step before considering fully turbulent trajectories.  

Unless the unsteady base flow is an exact solution to the NSE, care must be taken to ensure that the correct insights are extracted from the dataset.  This means that we should quantify the influence of any simulation error, averaging effects, or noise.  This can be achieved by varying the level of convergence or noise and assessing their effects.  In fact, the use of non-exact solutions should always be treated as a model beyond strict adherence to the dynamical systems theory.  Possible deviations from the theoretical descriptions should be kept in mind throughout the analysis process.

% ------------------------------ %  
\section{Approaches}
\label{sec:approaches}

{\color{black} In this section, we describe a series of data-driven and operator-based approaches for understanding perturbation dynamics about unsteady base flows. Table \ref{Table} briefly describes the main approaches, their typical inputs, and the computational considerations. When using data-driven approaches, flow features of interest must be present in the dataset, as these frameworks cannot infer characteristics or features beyond what is captured in the input data. In Table \ref{Table}, we discuss the operator-based methods assuming the operator is represented as a matrix, but matrix-free variants exist for nonmodal stability analysis \citep{Farrell1992} and resolvent analysis \citep{martini2020}. Matrix-based approaches tend to require constructing a large matrix and performing the singular value decomposition (SVD), whereas matrix-free approaches tend to involve iteratively solving forward and adjoint equations. The trade-off between the memory costs of constructing the matrix and the computational time required for forward and backwards solves should be weighed when deciding whether to take a matrix-based or matrix-free approach.}

\begin{table}
\caption{Summary of main techniques reviewed in this paper for studying flow dynamics around unsteady base states.}
\begin{center}
\footnotesize
\renewcommand{\arraystretch}{1.5}
\begin{tabular}{>{\centering\arraybackslash}m{45pt} m{55pt}  m{150pt} m{150pt}}
\br
Methods &  Inputs  & Descriptions &  Computational considerations \\
\mr
     POD, SPOD, DMD (\ref{sec:approaches_modal}) & Time-series data & Provides a set of modes to represent a dataset. The optimality of these modes depends on the formulation.  & Requires the SVD of a weighted data matrix. \\ 
     Nonmodal stability analysis (\ref{sec:approaches_nonmodal}) & Fundamental solution operator  & Provides the largest growth achievable by a perturbation and its spatial structure. & Requires construction and SVD of the fundamental solution operator at every time. \\
     OTD (\ref{sec:approaches_OTD}) & Linearized NSE  & Provides a set of orthogonal modes that spans the directions associated with the dominant growth of perturbations about a trajectory. & Requires solving for a set of modes forward in time. The selection of the initial OTD subspace is important. \\
     Resolvent variants (\ref{sec:approaches_resolvent}) & Resolvent operator & Provides forcing, response, and gains for a given variant of the resolvent operator. & Requires computing the SVD of the resolvent operator (often repeatedly for different frequencies).  \\
     Causality analysis (\ref{sec:approaches_causality}) & Time-series data & Identifies relationships of causality among different temporal signals over a given time window. & Requires considerable temporal resolution on time-series data, and may not converge with little data. \\
\br 
\end{tabular}
\end{center}
\label{Table}
\end{table}

\subsection{Data-driven modal decompositions}

\label{sec:approaches_modal}

Data-driven modal analyses \cite{HLBR-11,berkooz1993pod,kutz2016book,Schmid2022} can also be used on flows with a non-stationary base state, but careful consideration of the problem setup is necessary before proceeding. Two of the most widely used techniques for this purpose are proper orthogonal decomposition (POD) and dynamic mode decomposition (DMD). POD identifies dominant basis vectors that span the given data distribution in an optimal manner. In particular, this method finds a linear subspace which maximizes the captured variance in the dataset and provides the most compact and energy-efficient representation when the data distribution can be expressed as a linear subspace in which the covariances are zero (i.e., a hyperellipsoid). However, if the base flow evolves significantly in time, POD modes may struggle to properly capture the dominant dynamics of the system, as the base flow is no longer a fixed point in the state space \cite{Noack-03}.  Moreover, the mode deformation may play an important role as the base state evolves. Some noteworthy extensions of POD for the identification of the dominant dynamics about unsteady time-varying base flows include the conditional space-time POD developed for the prediction of intermittent or sudden burst events \cite{schmidt2019conditional}, the study of transient or non-stationary dynamics through the identification of space-time POD modes in a local temporal window \cite{frame2023space}, and the extraction of physically interpretable modes in near-periodic systems with the intrinsic phase-based POD (IPhaB POD) \cite{Borra2025IPhaB_POD}. Note that the incorporation of a windowing process, however, only proves effective for short time scales, and cannot provide a full characterization of an unsteady behavior over long time scales. {\color{black} A temporal proper orthogonal decomposition (TPOD) has also been developed to characterize the evolution of coherent structures across different flow regimes, including natural, controlled, and transitional states, offering a nuanced decomposition of time-varying base flows \cite{Gordeyev2013TSPOD}.}

An alternative approach that overcomes some of the limitations of standard POD analysis, particularly in handling the temporal nature of the base flow, is spectral proper orthogonal decomposition (SPOD) \cite{towne2018spectral,Schmidt2020Guide}. In contrast to traditional POD, SPOD operates in the frequency domain and identifies spatial modes associated with specific frequencies. This approach is especially useful for systems with a time-periodic or oscillatory behavior. Furthermore, recent work \cite{heidt2023spectral} extends SPOD to the analysis of harmonically forced flows (cyclostationary SPOD or CS-SPOD), enabling the distinction between time-periodic effects attributed to an external harmonic forcing and those associated with the time-periodic evolution of the base flow. In fact, this work establishes a direct connection between CS-SPOD and harmonic resolvent analysis (discussed in section~\ref{sec:harmonic}), with CS-SPOD providing a data-driven approach to studying frequency couplings and interactions.

Similarly, DMD offers a complementary approach to analyzing time-varying flows. In its original formulation \cite{schmid2008}, DMD identifies a set of non-orthogonal modes, each associated with a fixed complex frequency and growth/decay rate, to capture the dominant system dynamics in time. This decomposition is particularly informative in the cases where the temporal evolution of the base flow remains stationary or is time-periodic \cite{Schmid2022}. A similar challenge thus arises in the application of DMD to flows with time-varying base states. It is possible to circumvent this limitation in some instances by applying a temporal filter to isolate dynamical features that might present time-periodic features over a local temporal window. This approach is illustrated on the windowed variant of DMD for multi-resolution time-scale analysis described in \cite{kutz2016multiresolution}, and the online DMD for real-time tracking of the dominant dynamics in time-periodic systems in \cite{Zhang2019}. Alternatively, variational mode decomposition-based nonstationary coherent structure analysis for spatiotemporal data (VMD-NCS) analysis \cite{Ohmichi2024} provides a framework for identifying dynamically relevant coherent structures in nonstationary flows by decomposing the spatiotemporal field into intrinsic coherent structures that evolve in time, offering an alternative to traditional DMD when handling transient or intermittent dynamics about an unsteady base state.

In figure 2, we present three examples of modal analyses adapted to the study of time-varying flows, including space-time POD \cite{schmidt2019conditional}, cyclostationary SPOD \cite{heidt2023spectral}, and multi-resolution DMD \cite{kutz2016multiresolution}. The modal analyses presented thus far serve as powerful tools for analyzing the dynamics of the fluctuations about statistically-stationary base states.  While their extension to the study about unsteady base states is still in its early stages, further research and refinement will offer enhanced capturing of the time-varying characteristics of flows about unsteady base states.
% }

\begin{figure}
\centering {
\vspace{0cm}
{\hspace*{-0.2cm}\includegraphics[angle=0,width= 1.0\textwidth]{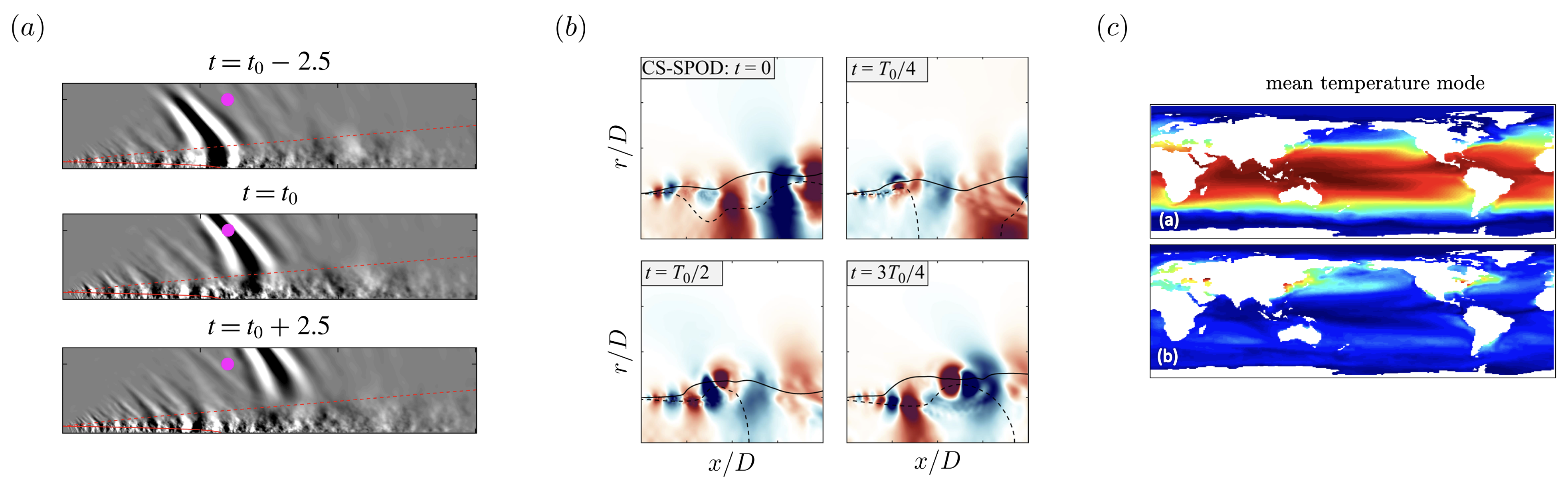}} }
\vspace{-0.5cm}
\caption{(a) Temporal evolution of the leading conditional space–time POD mode in a round supersonic jet \cite{schmidt2019conditional}, (b) real component of the pressure of the dominant CS-SPOD mode of a forced Mach 0.4 turbulent jet \cite{heidt2023spectral}, and (c) multi-resolution DMD mode of sea surface temperature data \cite{kutz2016multiresolution}. Reprint permissions granted by Cambridge University
Press and Society for Industrial and Applied Mathematics. 
}
\vspace{-0.cm}
\label{fig:modal_multigrid}
\end{figure}

\subsection{Nonmodal stability analysis}
\label{sec:approaches_nonmodal}

Nonmodal stability analysis can elucidate how perturbations grow when there are nonnormal eigenvectors for flows with time-varying base states \cite{Schmid2001,Schmid:2007}.  Here, we first discuss linear nonmodal approaches, and then briefly discuss nonlinear extensions. Often, when referring to nonmodal stability analysis, we seek optimal perturbations, which exhibit the largest transient growth through the linearized equations of motion. We can track these perturbations on top of a time-varying base flow, making them an important tool for studying unsteady flows.

To compute optimal perturbations, we consider the case of no forcing ($\check{\mathbf{f}}(t)=\mathbf{0}$), resulting in the 
linearized equations of motion 
\begin{equation} \label{eq:Linear}
    \dfrac{d\mathbf{q}'(t)}{dt}=\mathbf{L}[\bar{\mathbf{q}}(t)]\mathbf{q}'(t).
\end{equation}
To achieve no forcing, we consider cases where the base flow satisfies the Navier--Stokes equation, there is no perturbation forcing $\mathbf{f}'(t)$, and perturbations are small.
Under these conditions, the solution to \eqref{eq:Linear} is
\begin{equation}
    \mathbf{q}'(t)=\mathbf{A}(t)\mathbf{q}'(0).
\end{equation}
In the above equation, $\mathbf{A}(t)$ is the fundamental solution operator, which can be numerically approximated by
\begin{equation}
    \mathbf{A}(n \Delta t)=\prod_{j=1}^n e^{\mathbf{L}[\bar{\mathbf{q}}(j\Delta t)]\Delta t}.
\end{equation}
Given this fundamental solution operator, we compute the transient growth (or maximum possible amplification of initial energy density) 
\begin{equation} \label{eq:Growth}
    G(t)=\max_{\mathbf{q}'(0)\neq 0} \dfrac{||\mathbf{q}'(t)||^2_E}{||\mathbf{q}'(0)||^2_E}=\max_{\mathbf{q}'(0)\neq 0} \dfrac{||\mathbf{A}(t)\mathbf{q}'(0)||^2_E}{||\mathbf{q}'(0)||^2_E},
\end{equation}
where $||\cdot||_E$ is the energy norm, which depends on the flow of interest. For the case of the $L_2$-norm, the transient growth is simply the leading singular value $G(t)=\sigma_1$ of $\mathbf{A}(t)$, and the optimal perturbation is the leading right singular vector $\mathbf{v}_1$ given $\mathbf{A}(t)=\mathbf{U}\Sigma \mathbf{V}^{\rm T}$. 
This method shows, out of all possible perturbations, which will grow the largest at time $t$, and the shape of that initial perturbation. Notably, $G(t)$ does \emph{not} necessarily correspond to the same perturbation $\mathbf{q}'(0)$ for all $t$. This transient growth should instead be viewed as the energy envelope that no perturbation can cross. 

In many flows, this transient growth can be extremely large despite the eigenvalues containing negative real parts due to the nonnormality of the eigenvectors. This offers a linear route for the subcritical transition to turbulence \citep{Trefethen1993}, which could not be determined from standard linear stability analysis alone. Furthermore, transient growth can also come about from a time-varying flow, which also cannot be captured with standard linear stability analysis. 

\begin{figure}
    \colorbox{white}{\includegraphics[width=\textwidth,trim=2 2 2 2,clip]{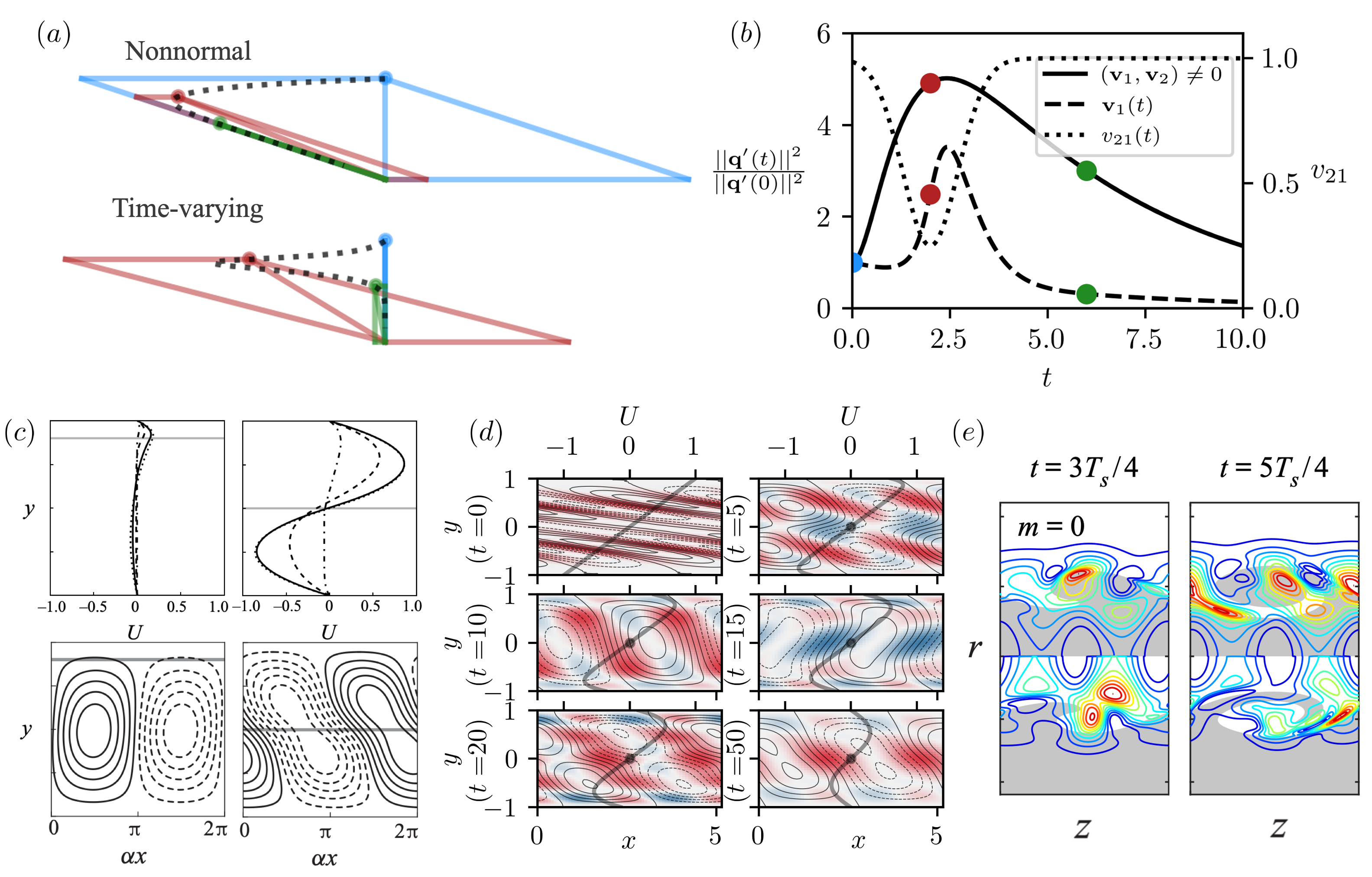}}
    \captionsetup[subfigure]{labelformat=empty}
    \begin{subfigure}[b]{0\textwidth}\caption{}\vspace{-10mm}\label{fig:Opta}\end{subfigure}
    \begin{subfigure}[b]{0\textwidth}\caption{}\vspace{-10mm}\label{fig:Optb}\end{subfigure}
    \begin{subfigure}[b]{0\textwidth}\caption{}\vspace{-10mm}\label{fig:Optc}\end{subfigure}
    \begin{subfigure}[b]{0\textwidth}\caption{}\vspace{-10mm}\label{fig:Optd}\end{subfigure}
    \begin{subfigure}[b]{0\textwidth}\caption{}\vspace{-10mm}\label{fig:Opte}\end{subfigure}
    \vspace{-5mm}

    \caption{(a) growth of an initial perturbation due to nonnormality and due to a time-varying eigenvector. The colors correspond to times in (b), the dot is the location of the perturbation $\mathbf{q}'$, and the other lines are the eigenvectors $y_1\mathbf{v}_1$, $y_2\mathbf{v}_2$. The black dotted line tracks the perturbation. (b) the energy of the perturbations in (a), and the $y$ value of the time-varying eigenvector. (c)-(e) example time-varying flows, and the evolution of their optimal perturbations for a particle-laden flow \citep{Chang_Chern_Chou_2021}, a decelerating channel \citep{Linot2024}, and a round jet undergoing a Kelvin–Helmholtz instability \citep{Nastro_Fontane_Joly_2022} (top panel is $\text{At}=0.25$, bottom panel is $\text{At}=-0.25$), respectively. Reprint permissions granted by Cambridge University Press.}
\label{fig:Opt}
\end{figure}

We illustrate these effects by considering how specific perturbations grow for a system with oblique and time-varying eigenvectors (i.e., we do not consider optimal perturbations here). First, we show the general equations for a diagonalizable time-varying linear system. Then we show an example of oblique, stationary eigenvectors, and an example of time-varying eigenvectors.

Consider the case where the linear operator in \eqref{eq:Linear} is diagonalizable such that
\begin{equation} \label{eq:eigen}
    \dfrac{d\mathbf{q}'}{dt}=\mathbf{V}(t)\Lambda(t)\mathbf{V}(t)^{-1}\mathbf{q}'.
\end{equation}
Letting $\mathbf{y}=\mathbf{V}(t)^{-1}\mathbf{q}'$, we can rewrite this equation in the eigenvector coordinate system as 
\begin{equation} \label{eq:coupled}
    \dfrac{d\mathbf{y}}{dt}=\left(\Lambda-\mathbf{V}^{-1}\dfrac{d\mathbf{V}}{dt}\right)\mathbf{y}.
\end{equation}
If the eigenvectors do not change in time, then the equations are decoupled and $\mathbf{y}$ grows or decays according to the eigenvalues. However, eigenvalues with a negative real part can still cause $\mathbf{q}'$ to grow when $\mathbf{V}$ is not an orthogonal matrix. We show this in two dimensions in figure\ \ref{fig:Opta} by plotting $y_1\mathbf{v}_1$, $y_2\mathbf{v}_2$, and $\mathbf{q}'$ for oblique eigenvectors and negative eigenvalues ($\lambda_1=-0.1$, $\lambda_2=-1.0$). Here, the content along each eigenvector decreases, but the perturbation still grows in size. This transient growth will be further amplified if the eigenvectors are made more oblique or the gap between eigenvalues increases.

Another contributor to transient growth is the time-varying nature of the eigenvectors.
As shown in \eqref{eq:coupled}, a time-varying eigenvector can lead to growth in $\mathbf{y}$ due to $\mathbf{V}^{-1}d{\mathbf{V}}/dt$. For example, if we consider only the case where $\mathbf{v}_1=[v_{11},v_{21}]^{\rm T}$ varies in time and $\mathbf{v}_2=[1,0]^{\rm T}$ for perturbations in $\mathbb{R}^2$ then
\begin{equation}
    \dfrac{dy_1}{dt}=\left(\lambda_1- \frac{\dot{v}_{21}}{v_{21}} \right)y_1.
\end{equation}
For real eigenvalues and eigenvectors, if $\lambda_1-\dot{v}_{21}/v_{21}>0$ then the content in this eigenvector will grow even if $\lambda_1<0$. Figures \ref{fig:Opta} and \ref{fig:Optb} highlight this transient growth due to a time-varying linear operator. Here we prescribe a specific Gaussian function for $v_{21}$, shown in figure \ref{fig:Optb}, and $v_{11}$ is chosen to make the eigenvector unit norm.
In this example, we again fix $\mathbf{v_2}=[1,0]^{\rm T}$ and we consider negative eigenvalues ($\lambda_1=-0.1$, $\lambda_2=-1$). Notably, the components of $\mathbf{y}$ are no longer decoupled, so the increase in $|y_1|$ induces an increase in $|y_2|$. When the linear operator is nonnormal or time-varying, computing optimal perturbations offers insight into the transient growth of the system that can not be determined from standard linear stability analysis. While the examples in figures \ref{fig:Opta} and \ref{fig:Optb} highlight the growth of a specific perturbation, recall that the optimal perturbation is the perturbation that grows the largest.

Optimal perturbations have predominantly been used to study transient growth in stationary base flows. For example, \citep{Butler1992} and \citep{Reddy1993} showed that predominantly spanwise streaks in plane Couette flow exhibit the largest transient growth and that this growth increases as $\text{Re}^2$. Similar results were discovered for plane Poiseuille \citep{Butler1992,Trefethen1993}, pipe flow \citep{Schmid1994}, and the Blasius boundary layer \citep{Butler1992}. Many of these early approaches could be formulated using the matrix exponential approach described above.  Unfortunately, the size of the fundamental solution operator $\mathbf{A}(t)$ depends on the spatial grid. This makes directly carrying out the SVD of $\mathbf{A}(t)$ computationally challenging for large grid sizes.

Alternatively, optimal perturbations can be computed using an adjoint method.  Instead of forming the entire fundamental solution operator $\mathbf{A}(t)$, the adjoint method allows us to compute the optimal perturbation by solving the linear equations forward and the adjoint equations backward in time repeatedly. The fundamental concept underlying the adjoint method is that the leading singular value of $\mathbf{A}(t)$ is equivalent to the square root of the largest eigenvalue of $\mathbf{A}^*(t)\mathbf{A}(t)$ \citep{barkley2008optimal}. Thus, repeatedly acting on some initial condition with the forward and adjoint equations is equivalent to power iterations of $\mathbf{A}^*(t)\mathbf{A}(t)$. This approach was first formulated by \citep{Farrell1992} for oceanic flows and was extended to investigate boundary layers \citep{Corbett2000,LUCHINI_2000}. The adjoint method can also be combined with windowing to seek perturbations initialized in a specific region of space \citep{MONOKROUSOS2010}. 

Although transient growth can be analyzed for time-varying base flows, fewer studies consider this case compared to steady base states. For example, periodic flows such as Stokes oscillatory flows \citep{Biau2016}, pulsatile pipe flows \citep{Xu2021,Moron2022}, periodic vortex shedding past a cylinder \citep{Abdessemed2009}, and flow past a compressor blade \citep{Mao_Zaki_Sherwin_Blackburn_2017} have been studied. Aperiodic flows have also been examined such as an impulsively stopped channel flow \cite{Nayak2017}, particle-laden flows \cite{Chang_Chern_Chou_2021}, time-dependent mixing layers \cite{Arratia_Caulfield_Chomaz_2013}, internal gravity waves \cite{Parker_Howland_Caulfield_Kerswell_2021}, Kelvin-Helmholtz billows \cite{Lopez-Zazueta_Fontane_Joly_2016}, variable-density jets \cite{Nastro_Fontane_Joly_2022}, and decelerating channel flow \cite{Linot2024}. In figure \ref{fig:Opt}, we highlight the unsteady base flow and optimal perturbations for a few of these studies.
Given the low number of studies on aperiodic flows, there is a strong need to develop libraries of canonical transient flow systems and begin investigating the transient growth of perturbations in these flows.

An important extension of linear nonmodal stability is nonlinear nonmodal stability. This is an analysis where the size of the initial perturbation is prescribed, and $G(t)=\max_{\mathbf{q}'(0)\neq 0} ||\mathbf{q}'(t)||^2_E/||\mathbf{q}'(0)||^2_E$ is sought for the fully nonlinear equations of motion \citep{Kerswell2018}. Care must be taken when performing nonlinear nonmodal stability because the optimization problem is nonconvex, so there is no guarantee that solution methods will converge to a global minimum. Some applications of this nonlinear method include boundary-layer flows \cite{Cherubini2010}, pipe flow \citep{Pringle2010}, and plane Couette flow \citep{Monokrousos2011}. Another recent extension of nonmodal stability analysis computes the mean energy amplification \citep{Frame2024} instead of the maximum amplification. If the optimal perturbations represent an unlikely perturbation, then looking at the mean amplification may be appropriate.

Thus far, we have discussed the time evolution of the single leading optimal perturbation. However, when the fundamental solution operator $\mathbf{A}(t)$ can be computed directly, the left singular vectors provide an orthonormal set of initial perturbations. Unfortunately, when we evolve this set of initial perturbations forward in time, they immediately lose orthogonality. In the next section, we discuss a method for tracking an orthogonal basis through time.

\subsection{Optimally time-dependent (OTD) mode analysis}
\label{sec:approaches_OTD}
\begin{figure}
    \centering
    \colorbox{white}{\includegraphics[width=1\linewidth]{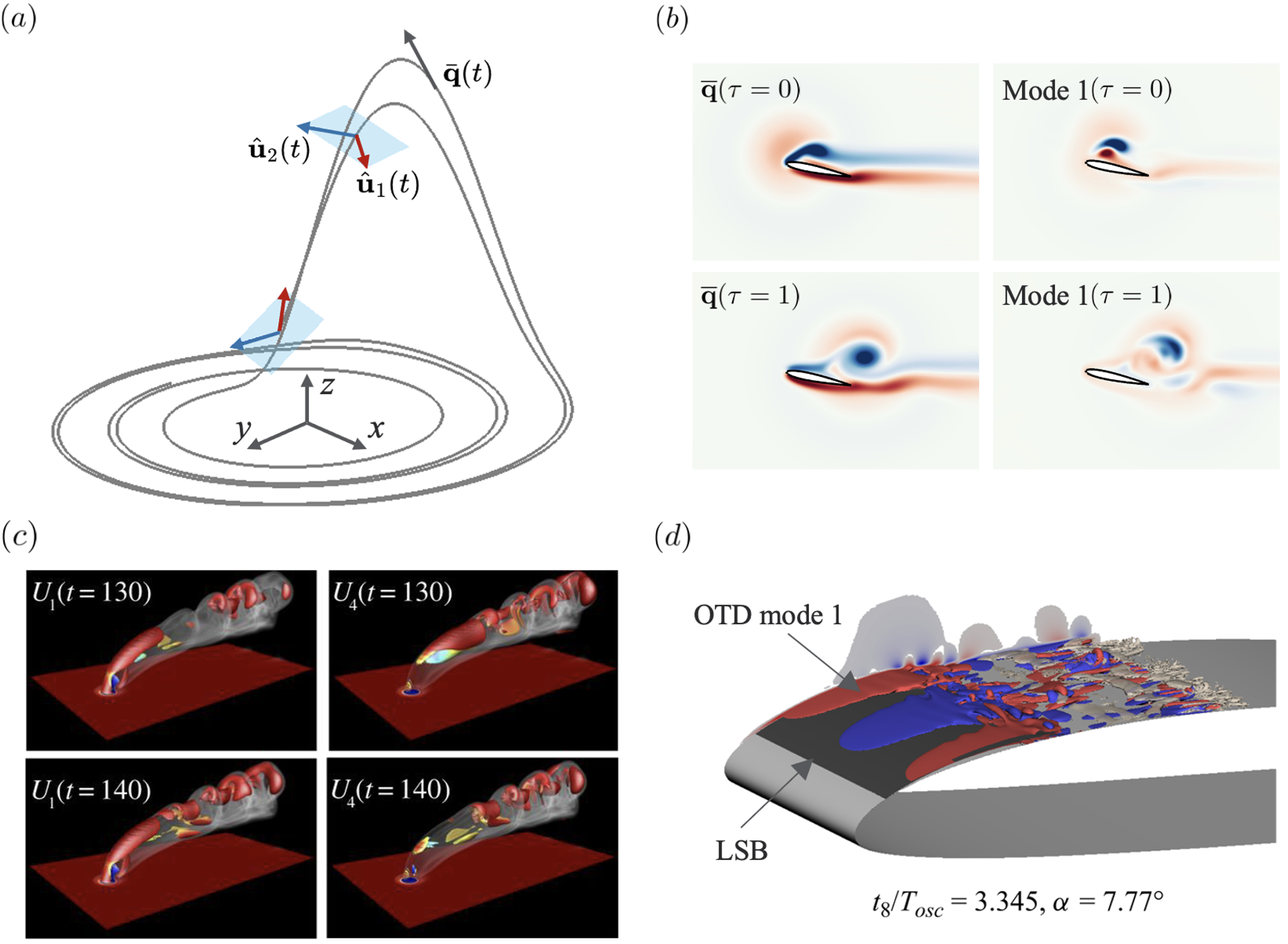}}
    \caption{(a)~The evolution of the base flow ${\bar{\mathbf q}}(t)$ and the optimally time-dependent (OTD) subspace spanned by the two most amplified direction ${\hat{\mathbf u}}_1(t)$ and ${\hat{\mathbf u}}_2(t)$ for an example of the R\"ossler system~\cite{zhong2025optimally}. (b)~Time-varying base flow and the leading OTD mode for vortex-airfoil interaction~\cite{zhong2025optimally}. (c)~The leading and fourth OTD mode for a vertical jet in cross-flow~\cite{babaee2016minimization}. (d)~The leading OTD mode as an indicator of the onset of absolute instability on a pitching airfoil~\cite{Kern_Henningson_2024}. Reprint permissions granted by Cambridge University Press and the Royal Society.}
    \label{fig:OTD_ver1}
\end{figure}

In this section, we introduce the optimally time-dependent (OTD) mode analysis, which focuses on the most amplified structures over a finite time interval with the orthogonal constraints on the modes. This method does not need the adjoint method to obtain initial perturbations. Instead, we choose an initial guess of the perturbations and evolve them {\color{black} forward with the} linearized NSE.
{\color{black}The} OTD mode analysis provides a framework for analyzing and understanding transient perturbation dynamics over unsteady base flows. 
The core idea of {\color{black}the} OTD mode analysis is to seek a set of time-evolving orthonormal bases that optimally capture the directions associated with the largest growth of perturbations about a trajectory (time-varying base flow). Unlike traditional modal decompositions about fixed base flows, {\color{black}the} OTD modes adapt to evolving dynamics.
In figure~\ref{fig:OTD_ver1}a, we show how {\color{black}the} OTD modes vary {\color{black}in} time for

a three-dimensional R\"ossler system, where the time-varying base flow is the trajectory ${\bar{\mathbf q}}(t)$. On top of the time-varying trajectory, the two most amplified OTD modes ${\hat{\mathbf u}}_1(t)$ and ${\hat{\mathbf u}}_2(t)$ span the OTD subspace (blue plane). 

Now{\color{black}, let us} derive the evolution equations for the OTD subspace~\cite{babaee2016minimization}. Consider a collection of $d$ perturbations $\mathbf{Q}^{\prime}(t)=[{\mathbf q}^{\prime}_{1}(t),{\mathbf q}^{\prime}_{2}(t),\dots,{\mathbf q}^{\prime}_{d}(t)]$, whose dynamics are determined by the linearized Navier--Stokes equations
\begin{equation}
    \frac{d \mathbf{Q}^{\prime}(t)}{dt}=\mathbf{L}[{\bar{\mathbf q}}(t)]\mathbf{Q}^{\prime}(t).  
    \label{eq:Lq}
\end{equation}
From this set of perturbations, we can {\color{black}consider} a low-rank approximation
\begin{equation}
    {\mathbf Q}^{\prime}(t)\approx\mathbf{U}_r(t)\mathbf{Y}_r(t)^{\rm T}
\end{equation}
with a set of $r$ orthonormal basis vectors
\begin{equation}
    \mathbf{U}_r(t)=[{\mathbf u}_{1}(t),{\mathbf u}_{2}(t),...,{\mathbf u}_{r}(t)]
\end{equation}
and their coefficients 
\begin{equation}
    {\mathbf Y}_r(t)=[{\mathbf y}_1(t), {\mathbf y}_2(t),...,{\mathbf y}_r(t)].
\end{equation}

The objective is to determine the rate of change of ${\mathbf Q}_r^{\prime}={\mathbf U}_r(t){\mathbf Y}_r(t)^{\rm T}$ that minimizes the Frobenius {\color{black}norm} to the linear evolution of {\color{black}${\mathbf Q}_r^{\prime}$} 
\begin{equation}\label{eq:var_prin}
F\left(\frac{d\mathbf{Y}_r}{dt},\frac{d\mathbf{U}_r}{dt}\right)=\bigg \| \frac{d(\mathbf{U}_r\mathbf{Y}_r^{\rm T})}{dt} -  \mathbf{L}\mathbf{U}_r\mathbf{Y}_r^{\rm T} \bigg \|_F,
\end{equation}
{\color{black} subject to the othonormality constraint ${\mathbf U}_r^{\rm T}{\mathbf U}_r={\mathbf I}$ (in the case of a Euclidean norm).}
This is essentially finding a set of OTD modes that best approximate the directions of maximum growth.  
By applying the orthogonal constraints {\color{black}to} the OTD modes, each mode stays linearly independent from {\color{black}the} other modes. 

{\color{black} We can solve~\eqref{eq:var_prin} with the orthonormality constraint by using Lagrange multipliers to turn this constrained optimization problem into an unconstrained optimization problem. 
In doing so, it is useful for us to take the time derivative of the orthonormal constraint, which yields 
\begin{equation}\label{eq:constraint}
    \langle \dot{\mathbf u}_{i}, {\mathbf u}_{j}\rangle+\langle {\mathbf u}_{i}, \dot{\mathbf u}_{j}\rangle=0,
\end{equation}
where $\left<\cdot,\cdot\right>$ is the inner product, and $\dot{\mathbf{u}}_i=d\mathbf{u}_i/dt$.
If we define $\bm{\phi}_{ij}=\langle {\mathbf u}_{i}, \dot{\mathbf u}_{j}\rangle$ ($\mathbf\Phi=[\bm{\phi}_{ij}]$), and observe that $\bm{\phi}_{ij}=-\bm{\phi}_{ji}$ from \eqref{eq:constraint}, then orthogonality is enforced when we minimize
\begin{equation}\label{eq:lag}
    \mathcal{G}\left(\frac{d\mathbf{Y}_r}{dt},\frac{d\mathbf{U}_r}{dt}\right)=\bigg \| \frac{d(\mathbf{U}_r\mathbf{Y}_r^{\rm T})}{dt} -  \mathbf{L}\mathbf{U}_r\mathbf{Y}_r^{\rm T} \bigg \|_F + \sum_{i=1}^{r}\sum_{j=1}^{r}\lambda_{ij}(\langle {\mathbf u}_{i}, \dot{\mathbf u}_{j}\rangle-{\bm \phi}_{ij}),
\end{equation}
where $\lambda_{ij}$ are the Lagrange multipliers.
The first-order optimality condition requires $\partial \mathcal{G}/\partial \dot{\mathbf{u}}_i=0$ and $\partial \mathcal{G}/\partial \dot{\mathbf{y}}_i=0$, which can be used to find
the evolution equations for ${\mathbf U}_r$ and ${\mathbf Y}_r$: 
\begin{equation}
        \frac{d{\mathbf U}_r}{d t}={\mathbf L}{\mathbf U}_r-{\mathbf U}_r({\mathbf U}_r^{\rm T}{\mathbf L}{\mathbf U}_r-\bm{\Phi}),
        \label{eq:Ur_Phi}
\end{equation}  
\begin{equation}
        \frac{d{\mathbf Y}_r^{\rm T}}{dt}=({\mathbf U}_r^{\rm T}{\mathbf L}{\mathbf U}_r-\bm{\Phi}){\mathbf Y}_r^{\rm T}.
    \label{eq:Yr_Phi}
\end{equation}
For a detailed derivation, we refer the reader to~\cite{babaee2019observation}. Although different choices of $\bm{\Phi}$ result in different OTD modes, the OTD subspace spanned by these modes is equivalent for any skew-symmetric choice of $\bm{\Phi}$~\cite{babaee2016minimization,blanchard2019analytical}.
Without loss of generality, $\mathbf{\Phi}=\mathbf{0}$ can be chosen for simplicity.
}

{\color{black} While the OTD subspace spanned by $\mathbf{U}_r$ is independent of $\mathbf{\Phi}$, there is no inherent ranking in the importance of these modes. We can define a unique, ranked set of OTD modes in its SVD form}
\begin{equation}
    {\mathbf U}_r{\mathbf Y}_r^{\rm T}=\hat{{\mathbf U}}_r(t)\bm\Sigma_r(t)\hat{{\mathbf Y}}_r(t)^{\rm T},
\end{equation}
where $\bm\Sigma_r (t) = {\rm{diag}}(\sigma_1(t),\sigma_2(t),...,\sigma_r(t))$.
The ranked spatial modes $\hat{{\mathbf U}}_r(t)=[\hat{\mathbf u}_{1}(t),\hat{\mathbf u}_{2}(t),...,\hat{\mathbf u}_{r}(t)]$
are used to characterize the most amplified directions of perturbations, and $\hat{{\mathbf Y}}_r(t)=[\hat{\mathbf y}_{1}(t),\hat{\mathbf y}_{2}(t),...,\hat{\mathbf y}_{r}(t)]$
is the coefficient matrix after rotation. Notably, the evolution of the OTD subspace ${\mathbf U}_r$ does \emph{not} depend on the coefficients ${\mathbf Y}_r$ in \eqref{eq:Ur_Phi}. It is only in rotating the OTD modes to find $\hat{{\mathbf U}}_r$ that we need to know the coefficients, which necessitates solving \eqref{eq:Yr_Phi}.

We can use the OTD subspace to define transient growth in a way similar to \eqref{eq:Growth}. Here, we define the maximum energy amplification of perturbation as
\begin{equation} \label{eq:GrowthOTD}
 g_1(t) = \frac{\| \mathbf q'_{*_1}(t)\|^2}{\|\mathbf q_{0_1}'(t) \|^2} =\frac{\sigma^2_1(t)}{\|\mathbf q_{0_1}'(t) \|^2}.   
\end{equation}
Where $\mathbf q_{*_1}'(t) = \sigma_1(t) \hat{\mathbf{u}}_1(t)$ is the most amplified perturbation and $\mathbf q_{0_1}'(t) =  \hat{\mathbf{U}}_r(t_0) \bm \Sigma_r(t_0) \hat{\mathbf Y}_r(t_0)^{\rm T} \hat{\mathbf y}_1(t)$ is
the optimal initial perturbation that leads to the maximum amplification at time $t$. If the optimal perturbations described in \ref{sec:approaches_nonmodal} lie within the initial OTD subspace, then the transient growth in \eqref{eq:Growth} and \eqref{eq:GrowthOTD} should be equivalent \cite{babaee2016minimization}.

In fluid mechanics, {\color{black} the} OTD mode analysis has been used to understand transient dynamics of various flows, including pulsating Poiseuille flow~\cite{kern2021transient}, the Blasius boundary layer~\cite{Beneitez_M_2023}, flow over a pitching airfoil~\cite{Kern_Henningson_2024}, and vortex-airfoil interactions~\cite{zhong2025optimally}. {\color{black} The} OTD mode analysis has also been used in designing control strategies for stabilizing unsteady flows~\cite{blanchard2019control,blanchard2019stabilization}.

Examples of {\color{black}the} OTD mode analysis on different time-varying systems are shown in figure~\ref{fig:OTD_ver1}. 
For a fluid system, such as the vortex-airfoil interaction shown in figure~\ref{fig:OTD_ver1}b, {\color{black}the} OTD modes identify time-varying structures where perturbations are amplified most about the unsteady base flow of vortex-airfoil interaction. As illustrated in figure~\ref{fig:OTD_ver1}c, the leading OTD mode ($U_1$) and a higher-order mode ($U_4$) capture different amplification mechanisms associated with the shear layer and the downstream vortical structures of a vertical jet. In addition, {\color{black}the} OTD modes are closely related to flow instability. In figure~\ref{fig:OTD_ver1}d, the location and the spanwise structure of OTD mode 1 indicates the onset of absolute instability, soon followed by rapid breakdown of the separation bubble on a pitching airfoil. The OTD mode analysis helps identify and track key dynamic features in unsteady flows by revealing when and where perturbations can be amplified.  Such insights can provide critical timing and location information for potential flow control applications.

The computational cost of solving for the OTD modes is roughly equivalent to solving the linearized equations \eqref{eq:Linear} forward $r$ times, which is much cheaper than solving forward all $d$ perturbations in \eqref{eq:Lq} since $r\ll d$. An adaptive algorithm has recently been developed to further reduce the computational cost by varying $r$ at each instance using the discrete empirical interpolation method~\cite{naderi2023adaptive}. Furthermore, in {\color{black}the} OTD mode analysis, there is no need to explicitly compute the matrix $\mathbf {L}[\bar{\mathbf{q}}(t)]$ if a linearized solver for \eqref{eq:Linear} is available. For a high-dimensional discrete fluid flow system, it is generally expensive to explicitly store the discrete linear operator $\mathbf {L}[\bar{\mathbf{q}}(t)]$ through a matrix-forming method.

Another important detail for performing {\color{black}the} OTD mode analysis is the selection of the initial conditions ${\mathbf{Q}'}$. Poor choices for initial conditions can lead to slow convergence or even divergence of the modes~\cite{babaee2016minimization}. Possible initial condition choices include random noise, POD modes, DMD modes, eigenvectors, and optimal perturbations.  Of these choices, random initialization is simple and widely used because it can be effective for exploring the solution space, but may require multiple runs to ensure convergence to a global optimum~\cite{zhong2025optimally}. In fluid systems, leveraging prior knowledge of the system to choose initial modes based on known flow patterns or instabilities can be effective in achieving fast convergence~\cite{kern2021transient}. Practitioners should carefully choose the initial conditions in {\color{black}the} OTD mode analysis to improve the accuracy and interpretability of the results.

An extension of the OTD framework can be considered with external forcing as in \eqref{eq:generalpert}.  This formulation leads to the forced optimally time-dependent (f-OTD) mode analysis, which is designed to study the impact of external forcing on transient amplifications in dynamical systems~\cite{donello2022computing}. 
The distinct utility of f-OTD lies in its ability to dissect how specific forcing inputs propagate through a system and identify critical response locations at each time instance. For a time-invariant base flow, the f-OTD modes converge towards the response modes obtained from resolvent analysis subjected to specific harmonic forcing{\color{black}s}~\cite{amirimargavi2023lowrank}. 
The f-OTD method is useful in extracting the time-dependent correlated structures for reactive flows~\cite{nouri2022skeletal, RAMEZANIAN2021113882} and two-dimensional decaying isotropic turbulence~\cite{amirimargavi2023lowrank}. The timing and location information revealed by {\color{black}the} f-OTD mode analysis is useful in designing real-time control strategies for systems with time-varying base flows.

\subsection{Time-varying input-output analysis} 
\label{sec:approaches_resolvent}

Next, let us present the input-output analysis for the linearized Navier--Stokes equations that are employed to identify and characterize primary linear 
energy-amplification mechanisms. 
In this context, we recast the linearized governing equations as a linear mapping between a forcing input and the flow states (response).  The forcing input term accounts for the nonlinear terms of the linearized equations and any given exogenous input.

The original formulation of the framework \cite{jovanovic2005componentwise,mckeon2010resolvent}, although limited to the analysis of statistically-stationary base flows, has proven to be informative in several scenarios. Some of these applications include the estimation of large flow scales structures in turbulent channel flows \cite{illingworth2018estimating}, the identification of a reduced-order model for supersonic turbulent boundary layers \cite{bae2020resolvent}, the characterization of coherent structures in turbulent pipe flows \cite{abreu2020spectral}, the prediction of scale interactions in sustaining wall-turbulence in \cite{mckeon2017engine}, the design of flow control strategies for separated flows over a NACA0012 airfoil \cite{yeh2019resolvent}, and modeling the effect of complex surfaces such as a turbulent channel flow with ribblets \cite{chavarin2019riblet}. For a comprehensive introduction to resolvent analysis in the context of fluid dynamics, including a review of its numerical implementation, a discussion on the interpretability of the energy norms, and several practical examples, refer to \cite{jovARFM21, Rolandi2024resolvent}. 

To derive the input-output form of the linearized governing equations, we first assume that the fluctuations about the base state are statistically stationary  such that the flow state $\mathbf{q}'(t)$ and forcing input $\check{\mathbf{f}}(t)$ in \eqref{eq:generalpert}, for a particular frequency $\omega$, can be rewritten as Fourier modes of the form
\begin{equation}
\label{eq:qFourier}    
    \mathbf{q}'(t) = \hat{\mathbf{q}} e^{\text{i}\omega t} \quad \text{and} \quad
    \check{\mathbf{f}}(t) = \hat{\mathbf{f}} e^{\text{i}\omega t},
\end{equation}
where $(\hat{\cdot})$ represents the Fourier mode. As we discussed in section~\ref{sec:preliminaries}, the spatial dependence of variables is omitted after discretization onto a finite grid or a set of basis functions. Following the same reasoning, the symbols $\hat{\mathbf{q}}$ and $\hat{\mathbf{f}}$ represent spatially discretized versions of $\hat{{q}}(\mathbf{x})$ and $\hat{{f}}(\mathbf{x})$, respectively. Substituting \eqref{eq:qFourier} in \eqref{eq:generalpert} gives
\begin{equation}
    \label{eq:govFourier}
    \text{i}\omega \hat{\mathbf{q}} = \mathbf{L}\hat{\mathbf{q}}+\hat{\mathbf{f}},
\end{equation}
which yields the input-output relationship between $\hat{\mathbf{f}}$ and $\hat{\mathbf{q}}$ as
\begin{equation}
    \label{eq:resolvent}
    \hat{\mathbf{q}} = \left(\text{i}\omega\mathbf{I} - \mathbf{L}\right)^{-1}\hat{\mathbf{f}}=\mathbf{\hat{H}}_{\omega}\hat{\mathbf{f}},
\end{equation} 
where the transfer function $\mathbf{\hat{H}}_{\omega}$ is the resolvent operator defined at a frequency $\omega$ \cite{trefethen2005spectra}. Note that in this context, the forcing term $\hat{\mathbf{f}}$ accounts for both an exogenous harmonic input and the small variations about the base flow of the nonlinear terms of the linearized governing equations \cite{Schmid2001}. However, if the linearization of the governing equations is performed about a solution of the Navier–Stokes equations, the nonlinear term can be neglected, and the forcing term would consist solely of $\mathbf{f'}$. 

\begin{figure}
\centering {
\vspace{0cm}
{\hspace*{-0.2cm}\includegraphics[angle=0,width= 1.0\textwidth]{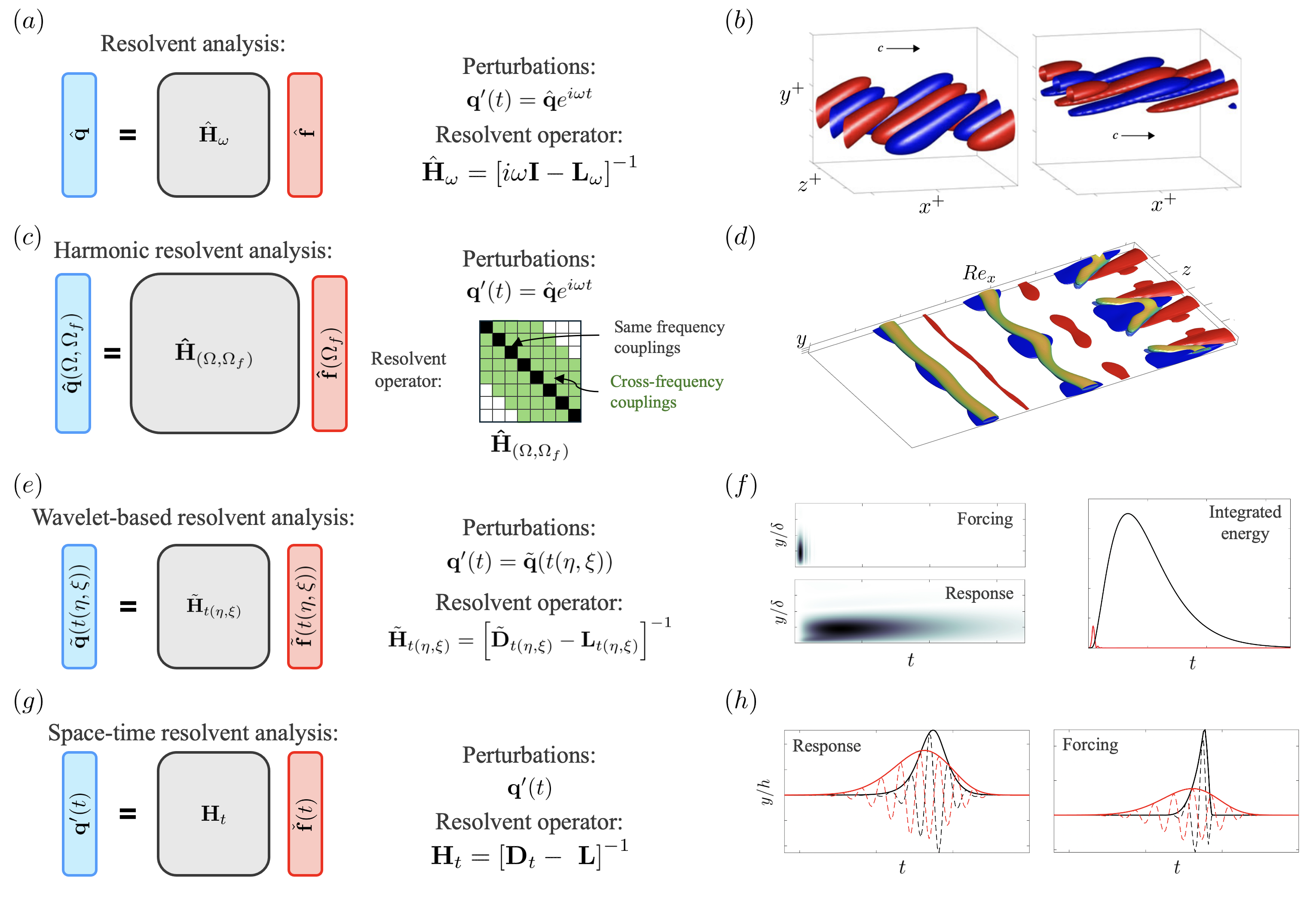}} }
\vspace{-0.3cm}
\caption{(a,c,e,g) Schematics of standard, harmonic, wavelet-based, and space-time resolvent analyses; (b) isosurfaces of streamwise velocity and wall-normal vorticity components of leading resolvent modes in a turbulent channel flow obtained from standard resolvent analysis \cite{zare2017colour}; (d) isosurfaces of streamwise harmonic response mode and of the second invariant of the velocity gradient tensor, colored by the vertical distance from the wall, in the controlled transition in a flat-plate boundary layer \cite{Rigas2021harmonic}; forcing (top) and response (bottom) wavelet resolvent modes in a turbulent channel and their corresponding amplitude over time (right) \cite{Ballouz2024growth}; (h) sparse (black) and non-sparse (red) response (left) and forcing (right) space-time resolvent modes in a turbulent channel flow with a sudden adverse pressure gradient \cite{lopezdoriga2024jfm}. Reprint permissions granted by Cambridge University Press and IOP Publishing. }
\vspace{0.cm}
\label{fig:multiGrid_resolvent}
\end{figure}

The coherent structures (in the form of Fourier modes) that are most amplified by the mean-linearized governing equations are determined via SVD of the resolvent operator \cite{mckeon2010resolvent,symon2018normal}. Note that this formulation only holds for flows that exhibit statistical stationarity in time. Several approaches in recent years have attempted to relax this assumption through different strategies, while leveraging the framework's capabilities. The following subsections present a brief compilation of these methods, including the harmonic resolvent analysis, wavelet-based resolvent analysis, and space-time resolvent analysis. In figure~\ref{fig:multiGrid_resolvent}, we present a collection of diagrams depicting each variant of resolvent analysis discussed in the following subsections, along with an illustrative example for every case.
% }

{\color{black} Let us briefly mention the data-driven formulation of resolvent analysis presented in \cite{Herrmann2021datadriven} for the transient evolution of linearly stable flows. This variant provides an equation-free formulation of resolvent analysis that relies on DMD to approximate the eigendecomposition of the underlying linear operator from data. The resulting data-driven resolvent operator is constructed by projecting the system dynamics onto the approximated eigenbasis. Though the computational cost of forming the resolvent operator is significantly reduced with this variant, it might fail to capture spatial and temporal features not included in the dataset that would otherwise be captured by the linearized equations. }

\subsubsection{Harmonic resolvent analysis}
\label{sec:harmonic}

Harmonic resolvent analysis identifies the energetically-dominant harmonic forcings and response modes that arise for time-periodic base flows. Assuming that the base flow has an oscillating-period $\tau$ (and a natural frequency $\omega=2\pi/\tau$), this formulation 
enables the study of triadic interactions between a harmonic forcing, harmonics of the base flow, and the corresponding output or response.   
In particular, since both the flow state $\mathbf{q}'(t)$ and the linearized operator $\mathbf{L}(t)$ are periodic, such that $\mathbf{L}(t)=\mathbf{L}(t+\tau)$, we can write them as Fourier series 
\begin{equation}
\label{eq:fourierHarm}
    \mathbf{L}(t)=\sum_{n=-\infty}^{\infty} \hat{\mathbf{L}}_ne^{\text{i}n\omega t}, \quad \text{and } \mathbf{q}(t)=\sum_{n=-\infty}^{\infty} \hat{\mathbf{q}}_ne^{\text{i}n\omega t},   
\end{equation}
where $(\cdot)_n$ denotes the $n$th harmonic of $\omega$. Substituting these expansions into \eqref{eq:generalpert} gives
\begin{equation}   
\label{eq:Texpansion}[\mathbf{T}\hat{\mathbf{q}}]_{k}=\text{i}k\omega\hat{\mathbf{q}}_{k}-\sum_{n=-\infty}^{\infty} \hat{\mathbf{L}}_{k-n}\hat{\mathbf{q}}_n=\hat{\mathbf{f}}_{k},
\end{equation}
where the operator $\mathbf{T}$ is a block-matrix operator of the form
\begin{equation}
\label{eq:Tharmonic}
    \mathbf{T}_{(\Omega_b,\Omega_f)} =
    \begin{bmatrix}
        \mathbf{\hat{H}}^{-1}_{-n} & & & & & & \\
         & \ddots & \vdots & \vdots & \vdots & \dots &\\
         & \dots & \mathbf{\hat{H}}^{-1}_{-1} & -\mathbf{\hat{L}}_{-1} & -\mathbf{\hat{L}}_{-2} &  \dots &\\
         & \dots & -\mathbf{\hat{L}}_{1} & \mathbf{\hat{H}}^{-1}_{0} & -\mathbf{\hat{L}}_{-1} &  \dots &\\
         & \dots & -\mathbf{\hat{L}}_{2} & -\mathbf{\hat{L}}_{1} & \mathbf{\hat{H}}^{-1}_{1} & \dots & \\
         &  & \vdots & \vdots & \vdots & \ddots & \\
         & & & & & & \mathbf{\hat{H}}^{-1}_{n}\\
    \end{bmatrix},
\end{equation}
and the subscripts $(\Omega_b,\Omega_f)$ denote the subsets of harmonic frequencies considered in the analysis. In particular, the subset $\Omega_b=\{-m\omega,...,m\omega \}$ with $m \in \mathbb{Z}$ denotes the harmonics of the periodic base flow, and $\Omega_f=\{-r\omega,...,r\omega \}$ with $r \in \mathbb{Z}$ represents the set of forcing frequencies. In this manner, the method accounts for the potential of the input frequency $k\omega$ to excite multiple harmonics given by the interaction of the periodic base state and $\mathbf{q}'$, and that are encapsulated by the coupling terms of the block-matrix operator. We define the diagonal elements 
\begin{equation}
    \mathbf{\hat{H}}^{-1}_{n\omega} = \text{i}n\omega\mathbf{I}- \hat{\mathbf{L}}_0,  
\end{equation}
which represent the linear response of the system at frequency $n\omega$ due to a forcing at the same frequency $\hat{\mathbf{f}}_n$.
The term $\hat{\mathbf{L}}_0$ represents the time-averaged periodic linear operator, corresponding to the zeroth harmonic of the Fourier expansion of the linear operator in \eqref{eq:fourierHarm}. The off-diagonal term $\mathbf{\hat{L}}_{k-n}$, on the other hand, encapsulates the coupling between a forcing frequency $k\omega$, and a harmonic of the base flow $n\omega$, resulting in a response at a frequency $(k-n)\omega$, as described by \eqref{eq:Texpansion}. These couplings arise due to the periodicity of the base flow, leading to fluctuations in the linearized operator $\mathbf{L}(t)$. As a result, the off-diagonal terms represent the influence of periodic fluctuations of the base flow in the linearized nonlinear terms, which induce frequency coupling and enable triadic interactions in the system.

Furthermore, the harmonic resolvent operator is defined as
\begin{equation}    \mathbf{\hat{H}}_{(\Omega_b,\Omega_f)}=\mathbf{T}^{-1}_{(\Omega_b,\Omega_f)},
\end{equation}
and the input-output formulation of the harmonic resolvent framework is given by
\begin{equation}
\label{eq:harmonicR}
    \mathbf{\hat{q}}(\Omega_b,\Omega_f) = 
    \mathbf{\hat{H}}_{(\Omega_b,\Omega_f)}
    \mathbf{\hat{f}}(\Omega_f),
\end{equation}
where $\mathbf{\hat{q}}(\Omega_b,\Omega_f)$ and $\mathbf{\hat{f}}(\Omega_f)$ represent the state response and the forcing term across all harmonics of the base flow $\Omega$ and forcing frequencies $\Omega_f$ in the form of Fourier modes.

This approach is utilized in \cite{harmonic2020padovan} to identify harmonic interactions between harmonic forcings and the base flow of an airfoil under near-stall conditions. A similar procedure is described in \cite{Rigas2021harmonic} to study and characterize the influence of a spanwise harmonic forcing on a flat-plate boundary layer. Other noteworthy implementations of this analysis include the study of subharmonic perturbations on a forced incompressible axisymmetric jet in \cite{padovan2022analysis}, the sensitivity analysis of a turbulent separation bubble to actuation at different frequencies in \cite{wu2022response}, and the identification of cross-frequency interactions in cavity flows  \cite{islam2024identification}. Moreover, \cite{farghadan2024efficient} provides an enhanced formulation of this method with higher numerical efficiency based on a time-stepping approach, along with the application of this approach to characterize the harmonic interactions between the periodic motions observed in flow over an airfoil at low Reynolds numbers. 

\subsubsection{Wavelet-based resolvent analysis}

Wavelet-based resolvent analysis incorporates into its formulation a wavelet transform \cite{meyer1992wavelets} in the temporal direction \cite{ballouz2023wavelet,ballouz2024wavelet}. Wavelets are wavelike basis functions with a high degree of spatial and temporal localization. 
Additional details regarding the properties of wavelet functions can be found in section \ref{sec:wavelets}. The incorporation of wavelet functions to resolvent analysis enables the identification of dominant dynamics without restrictions in the temporal characteristics of the base flow, allowing for the identification of energy amplification mechanisms about both statistically-stationary and time-varying base flows. In this context, the following forms for the state response and the forcing are proposed
\begin{equation}
\label{eq:qWavelet} 
    \mathbf{q}'(t) = \tilde{\mathbf{q}}(t(\eta,\xi)) \quad \text{and} \quad 
    \check{\mathbf{f}}(t) = \tilde{\mathbf{f}}(t(\eta,\xi)) ,
\end{equation}
where we use the symbol ($\tilde{\cdot}$) to denote a wavelet mode, and the parameters $(\eta,\xi)$ represent temporal translation and scaling of the wavelet mode. In practice, this wavelet transform is applied via the discrete wavelet transform matrix $\mathbf{W}_{t(\eta,\xi)}$, in which the wavelet parameters $(\eta,\xi)$ are embedded. The wavelet-based resolvent operator is thus defined as
\begin{equation}
    \label{eq:resolventWavelet}
    \tilde{\mathbf{q}}(t(\eta,\xi)) = (\tilde{\mathbf{D}}_{t(\eta,\xi)} - \mathbf{L})^{-1}\tilde{\mathbf{f}}(t(\eta,\xi))= \tilde{\mathbf{H}}_{t(\eta,\xi))}\tilde{\mathbf{f}}(t(\eta,\xi)),
\end{equation}
where the term ${i}\omega\mathbf{I}$ in~\eqref{eq:resolvent} is now substituted by a wavelet-transformed time differentiation matrix $\tilde{\mathbf{D}}_{t(\eta,\xi)}$. The localization achieved by the incorporation of the wavelet transform in the temporal direction enables the identification of transient, or simply time-localized, linear amplification mechanisms in systems with a time-varying base flow. 

In this context, the choice of wavelet basis should be tailored to the system of interest. For instance, \cite{ballouz2023wavelet,ballouz2024wavelet} use Shannon and Daubechies-16 wavelets \cite{mallat2001wavelet,najmi2012wavelets}, since Shannon wavelets act like a bandpass filter in the frequency domain and provide a sharp frequency separation, and Daubechies wavelets also function as a bandpass filter but offer higher temporal localization (though at the expense of a denser time-differentiation operator). The relevance of the structure of the time-differentiation operator $\tilde{\mathbf{D}}_{t(\eta,\xi)}$ should be highlighted as well. While highly sparse finite-difference schemes may be sufficient to retain the main spatiotemporal features of some systems, denser schemes with a higher computational cost are needed for the analysis of systems with a high number of degrees of freedom. 

The work in \cite{ballouz2023wavelet,ballouz2024wavelet} showcases the implementation of this approach on a turbulent Stokes boundary layer (time-periodic base flow) and on a turbulent channel flow with a sudden lateral pressure gradient (transient base flow). Further implementations of this work include the study of the effect of a temporally-localized forcing mode (identified by wavelet-based resolvent analysis) on the transient mechanisms observed in a turbulent minimal flow unit in \cite{Ballouz2024growth,ballouz2025transient}. 

\subsubsection{Space-time resolvent analysis}

A generalized space-time formulation of resolvent analysis can study base flows without restrictions on their temporal nature. 
The input-output formulation of the space-time resolvent framework is given by
\begin{equation}
    \label{eq:resolventSpaceTime}
    \mathbf{q}'(t) = \left(\mathbf{D}_t - {\mathbf{L}}\right)^{-1} \check{\mathbf{f}}(t)= {\mathbf{H}}_{t}\check{\mathbf{f}}(t),
\end{equation}
where $\mathbf{D}_t$ now represents a generalized finite-difference operator.  
The principal advantage of the space-time analysis resides in the lack of restrictions enforced on the temporal nature of the spatiotemporal coherent structures. Its main drawback, however, is the increased computational expense of this approach compared to other variants of resolvent analysis. 
This formulation is exploited in \cite{ldoriga2023sparse,lopezdoriga2024jfm} using a second-order finite difference operator in time to analyze a turbulent Stokes boundary layer (time-periodic base flow) and a turbulent channel flow with a sudden adverse pressure gradient (transient base flow). In addition, \cite{ldoriga2023sparse,lopezdoriga2024jfm} combine the space-time analysis with a sparse formulation of resolvent analysis by introducing an $L_1$-norm term to the optimization problem posed by the SVD. This approach allows for the identification of spatiotemporal coherent structures with a great degree of localization or sparsity, with the added flexibility of externally adjusting the degree of spatiotemporal localization. Such flexibility has potential implications for control systems and optimal actuator placement. A similar motivation underpins the work in \cite{skene2022sparseForcing}, where the authors identify spatially-localized structures that can trigger amplified responses in the context of localized flow control actuation. 

Despite the advancements in the extensions of resolvent analysis developed in an effort to capture the time-varying nature of the base flow, the resulting computational cost must be carefully considered. For instance, assuming that the discrete standard resolvent operator in \eqref{eq:resolvent} has dimensions $(N \times N)$, the harmonic resolvent operator in \eqref{eq:harmonicR} that considers the forcing frequencies $\Omega_f$ has dimensions $(N N_{\Omega_f} \times N N_{\Omega_f})$, where the factor $N_{\Omega_f}$ represents the number of harmonics contained in the set $\Omega_f$. Moreover, both wavelet-based (see \ref{eq:resolventWavelet}) and space-time (see \ref{eq:resolventSpaceTime}) formulations of the resolvent have dimensions $(N N_t \times N N_t)$ with $N_t$ denoting the number of collocation points in the temporal direction. As a result, these formulations require increased computational resources, with higher dimensionality potentially making the analysis prohibitive, particularly when analyzing systems characterized by fine timescales or temporally localized events. Recent efforts to improve the computational efficiency of resolvent analysis-based methods involve the use of random projections \cite{moarref2013channels,ribeiro202randomized}, timestepping methods \cite{martini:2020,towne2022efficient}, or both simultaneously \cite{farghadan2024scalable,farghadan2024efficient}. The incorporation of these tools may alleviate some of the increased computational cost introduced by the discretization of the temporal direction in the extended resolvent-based frameworks. 

\subsection{Causality analysis}
\label{sec:approaches_causality}

In the realm of data-driven methods, several efforts have attempted to capture the principal features that drive and sustain fluid flows, usually at the expense of a large compilation of high-fidelity datasets. 
Here, we provide an overview of causal inference methods (also referred to as causality analyses) for fluid dynamics. The notion of causality, first introduced by Wiener \cite{wiener1956theory}, identifies measurable relationships of cause and effect between two events, where the former leads to the latter. 
In this context, the identification of relationships of cause and effect between the variables of the system, which helps uncover the dominant dynamics about base flows, with no restrictions on their temporal nature. This has direct implications for the study of unsteady flows, as causal inference methods offer an alternative approach to understanding the primary mechanisms that govern their dynamics. A key aspect of causality analyses is their ability to infer the directionality of the interactions, identifying the variables that actively drive the temporal evolution of others. 

\begin{figure}
\centering {
\vspace{0cm}
{\hspace*{0cm}\includegraphics[angle=0,width= 1\textwidth]{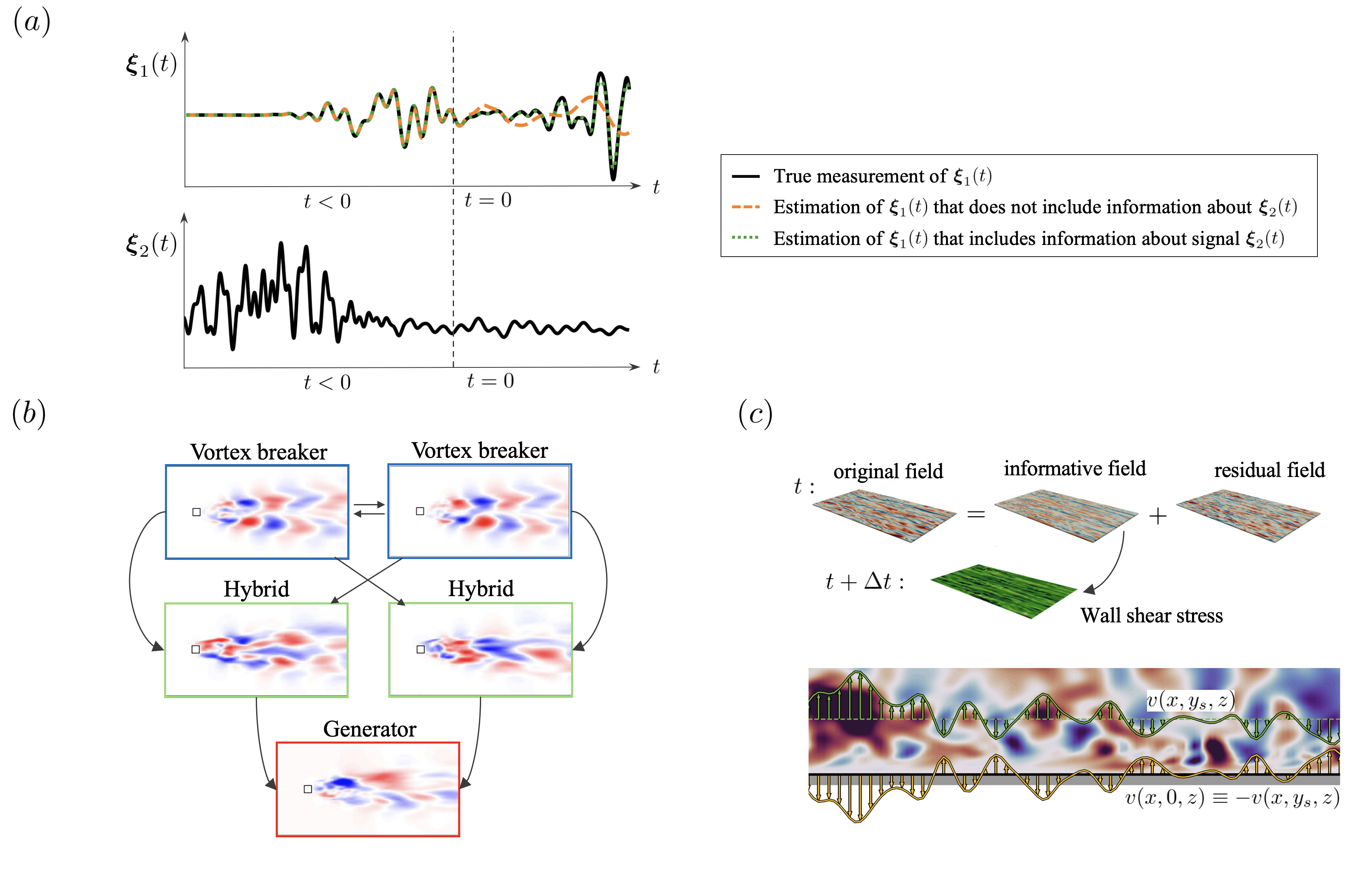}} }
\vspace{-0.3cm}
\caption{(a) Temporal evolutions of two signals $\boldsymbol{\xi}_1(t)$ and $\boldsymbol{\xi}_2(t)$, and the influence of previous information about $\boldsymbol{\xi}_2$ on estimation of the future state of $\boldsymbol{\xi}_1$ (assuming that there is in fact causation from $\boldsymbol{\xi}_2$ to $\boldsymbol{\xi}_1$); 
(b) primary causal interactions between the large-scale coherent structures observed in flow past a square cylinder \cite{martinez2022causality}; (c) diagram of causal inference model based on a decomposition of the flow field on its informative and residual components (top), and implementation of this decomposition for the prediction of wall shear stresses and the development of an opposition control for drag reduction in a turbulent channel (bottom) \cite{arranz2024informative}. Reprint permissions granted by Cambridge University Press.
}
\vspace{-0.3cm}
\label{fig:multiGrid_causality}
\end{figure}

Here, we present two families of causal inference methods: those based on Granger's concept of causality \cite{granger1969}, and those based on information theory and the concept of Shannon entropy \cite{shannon1948}. A diagram that further illustrates the notion of causality, along with two representative examples of causality analyses, is shown in figure~\ref{fig:multiGrid_causality}. First, figure~\ref{fig:multiGrid_causality}(a) depicts the temporal evolution of two signals $\boldsymbol{\xi}_1(t)$ and $\boldsymbol{\xi}_2(t)$, and how knowledge of $\boldsymbol{\xi}_2$ improves the estimation of $\boldsymbol{\xi}_1$ when there is a relationship of causality from $\boldsymbol{\xi}_2$ towards $\boldsymbol{\xi}_1$. In addition, figure~\ref{fig:multiGrid_causality}(b) depicts the main causal interactions between the POD modes behind a square cylinder identified in \cite{martinez2022causality}, and figure~\ref{fig:multiGrid_causality}(c) represents a schematic of the decomposition of the flow field based on regions of informative and residual components (top), used to predict the wall shear stress and to develop the opposition control for drag reduction in a turbulent channel (bottom) described in \cite{arranz2024informative}.

Causal inference methods based on Granger's concept of causality build upon the statistical approach \cite{granger1969}. In this framework, causality is measured as the improvement in the prediction of a temporal variable $\boldsymbol{\xi}_1(t)\in\mathbb{R}^n$, before and after incorporating past values of another variable $\boldsymbol{\xi}_2(t)\in\mathbb{R}^n$. 
Granger's original formulation employs a first-order regression on the available dataset, thereby restricting causal relationships to linear interactions between pairs of temporal variables. In this context, causality is measured by comparing the prediction errors of two regressions: one that includes $\boldsymbol{\xi}_2(t)$ and one that does not. If the regression that incorporates $\boldsymbol{\xi}_2(t)$ shows a significant improvement in predictive accuracy, the method infers a causal relationship from $\boldsymbol{\xi}_2(t)$ to $\boldsymbol{\xi}_1(t)$. Since Granger's approach assumes that a variable's past influences its future state only through external variables, it does not account for the possibility that a variable's own past values may affect its future state (self-induced causality). This formulation is extended to the identification of $n$-order interactions in \cite{ldoriga2024granger} by incorporating higher-order terms into the estimation, enabling the identification of causal interactions between the POD coefficients of the instantaneous velocity components in turbulent square duct flows.

Causal inference methods based on information theory hinge on the concept of Shannon entropy \cite{shannon1948}, which quantifies the uncertainty or randomness in a system. These approaches measure how much the uncertainty (entropy) of one signal is reduced by knowing the past states of another signal(s), thus inferring directional causal dependencies. Unlike Granger causality, information-theoretic methods do not impose restrictions on the nature of the causal interactions, allowing for a broader range of relationships to be identified. Originally, and similarly to Granger-based causality, these methods were limited to pairwise interactions, that is, causal relationships between a pair of signals, as in the study of causal relationships between the mean shear stresses and streamwise streaks and rolls observed in the logarithmic layer of a minimal turbulent channel flow \cite{bae2018causality}, the causal interactions between energy-eddies of different spatial scales identified in the buffer and logarithmic layers \cite{lozano2020causality}, and the identification of the coherent structures responsible for the generation of large-scale structures behind a square cylinder \cite{martinez2022causality}. Recent advancements have extended the analysis from pairwise interactions to multi-variable relationships of causality, and have also accounted for self-induced interactions (that is, a given signal influencing its own future state). These extensions have been applied to study the turbulent cascade in \cite{lduran2022information,MartinezSanchez2024causality}, the identification of causal interactions between streaks and bursts in wall-bounded turbulence \cite{Ling2024causality}, and the development of models to predict shear stresses, which are later incorporated in the design of an instantaneous opposition control to reduce drag in a turbulent channel \cite{arranz2024informative}.

The two families of causality methods presented are fundamentally different. Information-theoretic methods, while not restricted to any specific form of causality, do not directly reveal the nature of interactions, whereas Granger-based methods allow for the direct inference of the order of the interactions. Granger analyses generally require less data and are computationally less expensive compared to information theory-based methods, which need larger datasets for convergence. In addition, Granger-based methods do not account for either self-induced (the past of one variable conditioning its own future state) interactions, while some information theory-based methods do. Nonetheless, the direct connection between the linear Granger causality and DMD described in \cite{gunjal2023granger} can be informative in some instances. We also acknowledge the inherent limitations of causal analyses. In particular, the formulations of these methods are usually reliant on the observability of enough relevant states or variables, which may not always be available. In addition, the performance of certain causal analyses may suffer in the presence of stochastic noise, which can obscure relevant causal interactions in some instances \cite{MartinezSanchez2024causality}. Nonetheless, we highlight that both families of causal inference methods impose no restrictions on the temporal evolution of the base flow, which renders them well-equipped for the study of dominant dynamics about both statistically-stationary and time-varying (periodic and aperiodic) base flows. Other contributions focused on causal inference that are not based on the families of causality frameworks described in this section, include the work in \cite{Encinar2023isotropic,Encinar2024growth}. 

\subsection{Other Approaches}
\label{sec:approaches_other}

\subsubsection{Wavelet analysis}
\label{sec:wavelets}

Another method is the wavelet analysis, which, in the context of time-varying base flows, can identify and characterize the dominant dynamics across different spatial and temporal scales. Due to the localized nature of the wavelets, they effectively capture transient and intermittent phenomena, as well as sharp gradients, with sensitivity depending on the choice of wavelet basis \cite{daubechies1990wavelet}. Moreover, the intrinsic multi-scale resolution of the wavelet provides a localized basis for different temporal scales with equal resolution for all scales. One particularly relevant property of wavelets for the study of the dynamics about time-varying base flows is that all wavelet functions oscillate about zero, 
which makes them bandpass filters. This allows them to identify sudden fluctuations or non-periodic trends of the base flow. For a detailed discussion on wavelet theory and its applications, see \cite{daubechies1992ten}.

In the field of fluid dynamics, several approaches have resorted to the use of wavelets, but their implementation is generally restricted to the spatial directions. Recently, however, there has been an effort to incorporate the wavelet analysis into the temporal direction. Some of these contributions include the study of transient and intermittent phenomena in turbulent wakes and boundary layer flows in \cite{rinoshika2020application}, the multi-scale wavelet-based analysis of high-speed compressible turbulent boundary layers in \cite{khujadze2022wavelet}, the wavelet-based POD analysis to predict burst events in a forced flow over a 2-torus described in \cite{madhusudanan2025predicting}, and the previously discussed wavelet-based variant of resolvent analysis introduced in \cite{ballouz2023wavelet, ballouz2024wavelet}. 

\subsubsection{Network-Based Techniques}
Network-based techniques, which combine concepts from network science and dynamical systems theory, specialize in analyzing connections and interactions over a group of elements \cite{Newman18,Newman:SIAMReview03,Dorogovtsev10,Estrada12}.  Analysis of unsteady vortical flows is a key candidate to benefit from network science, because vortical structures are influenced by the rich interactions among them over a range of scales \cite{Taira:PAS22, Iacobello:PA21}.  Such vortical interactions can, in fact, be studied by quantifying the induced velocity \cite{Nair:JFM15, Taira:JFM16, Murali:JFM21} and the energy transfer \cite{Nair:PRE18}.  With network-based analysis tools, dominant interactions can be extracted from unsteady flows to model and control their dynamical behavior \cite{Nair:PRE18, Murali:PRE18, Murayama:PRE18, Murali:JFM21}. 

For a flow with an unsteady base state, a time-varying network can be used to quantify dynamical interactions. For instance, perturbation dynamics on top of a turbulent flow can be studied with the network broadcast analysis \cite{Yeh:JFM21}.  This approach reveals the most amplified disturbances broadcast over the time-evolving turbulence network. In fact, the broadcast analysis shares similarities in its formulation with the aforementioned resolvent analysis. Network broadcast analysis has identified flow structures that can optimally modify turbulent flows, as demonstrated in cases of two-dimensional isotropic turbulence. This suggests its potential for guiding time-adaptive flow control efforts in unsteady flows.

\subsubsection{Hilbert-Huang Transform}
Spectral analysis of a transient signal cannot be performed easily with the Fourier transform due to its foundation on sinusoidal (periodic) basis functions \cite{Bracewell99}. The Hilbert transform can address this issue and determine the instantaneous frequency of signals, but it assumes that the signal is statistically stationary. A non-stationary signal can be analyzed by combining the empirical mode decomposition (EMD) with the Hilbert transform. 

The EMD decomposes a given signal into a set of intrinsic mode functions (IMFs) through a process called sifting. This process fits a cubic spline through all local maxima of the signal and another cubic spline through all the local minima. The average of the two splines is then subtracted from the original signal, and the difference constitutes the IMF. This process is continued until all IMFs capture the fluctuations contained in the original signal. This procedure empirically determines the temporal change in the signal, generally from small to large-scale fluctuations. These IMFs can be used to form a model for the time-varying base state discussed earlier in this paper. Once the IMFs are identified and used to remove the non-stationary component of the signal, the Hilbert transform can then be applied to the IMFs to identify instantaneous frequency components and the associated structures. The overall approach that combines Hilbert transform and EMD to analyze non-stationary signals is called the Hilbert-Huang transform (HHT) \cite{Huang:1998}.

The HHT was originally conceived to analyze one-dimensional signals. Since then, the analysis has been extended to multiple dimensions. Because the extension beyond one dimension is computationally taxing \cite{Thirumalaisamy:IEEESPL18}, there have been limited applications in fluid dynamics. However, some recent studies have used HHT to study two-dimensional flows obtained from experiments \cite{Ansell:AIAAJ20} and simulations \cite{deSouza:TCFD24}. This analysis technique (along with EMD and the Hilbert transform) may help future analysis of transient flow physics and stimulate new approaches.

% ------------------------------ % 
\section{Concluding Remarks}
\label{sec:discussion}

Studying the dynamics about time-varying base flows is important to advance our understanding of many fluid systems. However, the approaches for studying these flows are not as well developed as for time-invariant or periodic base flows. Unsteady base flows present a unique challenge because even defining the base flow and the perturbation may present difficulties. 

Here, we presented promising emergent techniques to approach unsteady flows and discussed some of the applications reported in the literature. First, we considered data-driven modal techniques. Although many of these methods do not account for aperiodic, transient dynamics, conditional SPOD, and online DMD are possible candidates. Then, we considered operator-based techniques. This includes nonmodal stability analysis, optimally time-dependent mode analysis, and time-varying input-output analysis. Nonmodal stability provides a natural extension to standard linear stability analysis for time-varying systems by identifying the optimal perturbations. OTD shares similarities to optimal perturbations by identifying an orthonormal basis that identifies the directions of growth when following a trajectory. The time-varying input-output methods that can be used with time histories of aperiodic, unsteady base flows include wavelet-based resolvent and space-time resolvent. 

We also discussed causality analysis for understanding the causal relationship of time series about unsteady flows. In particular, we highlight approaches based on transfer entropy and Granger analysis. 
Finally, we briefly presented wavelet analysis, network-based techniques, and HHT for studying transient flow physics. Glaring omissions in this paper are deep learning techniques \cite{Brunton2020} such as autoencoder-based compression \cite{Hinton2006,Milano2002,Murata2020,Floryan2022,Smith2024homology}, surrogate models \citep{Vlachas2018,Pathak2018a,Linot2020}, physics-informed neural networks \citep{Raissi2019}, and super-resolution \cite{Fukami2019,Fukami2020}, whose coverage would require an equally long -- if not longer -- presentation of the materials.  These types of approaches may also advance our understanding of unsteady flows. 

We hope that this discussion on existing methods for extracting dominant dynamics about unsteady base flows both motivates the widespread use of these methods and inspires further development into new methods. Transient flows are common and often behave differently from steady flows, making this an important area of future research.

% ------------------------------ % 
\section*{Acknowledgments}

KT thanks Prof.~Genta Kawahara for the invitation to submit this perspective paper to Fluid Dynamics Research.  This paper is partly based on a presentation given at the 26th International Conference of the Theoretical and Applied Mechanics (ICTAM 2024) by KT.  We are thankful to Laura Victoria Rolandi and Jonathan Tran for their feedback on our manuscript.  We acknowledge the support from the US Air Force Office of Scientific Research (grant number: FA9550-21-1-0178), the US Army Research Office (grant number: W911NF-21-1-0060), and the US Department of Defense Vannevar Bush Faculty Fellowship (grant number: N00014-22-1-2798).

% ------------------------------ % 
\bibliographystyle{unsrt}
\bibliography{Master,Floquet,Taira_all,Taira_all2}

\end{document}